\documentclass[12pt]{article}

\usepackage{amsmath}
\usepackage{latexsym}
\newcommand\be{\begin{equation}}
\newcommand\ee{\end{equation}}
\newcommand\bea{\begin{eqnarray}}
\newcommand\eea{\end{eqnarray}}
\newcommand\beas{\begin{eqnarray*}}
\newcommand\eeas{\end{eqnarray*}}

\def\tr{{\rm Tr}}

\begin{document}
\title{Matrix Model Maps in AdS/CFT\footnote{Brown- HET-1452; WITS-CTP-023}}

\author{Aristomenis Donos and
Antal Jevicki\
\\ Physics Department\\
Brown University\\
Providence, Rhode Island 02912, USA \\
\and
Jo\~ao P. Rodrigues \\
School of Physics and Centre for Theoretical Physics \\
University of the Witwatersrand\\
Wits 2050, South Africa\\
}

%\begin{document}
\maketitle

\begin{abstract}
We discuss an extension of a  map between between BPS states  
and free fermions. The extension involves states associated with a full two matrix problem 
which are constructed using a sequence of integral equations. A two parameter set of
matrix model eigenstates is then related to states in SUGRA. Their wavefunctions are characterized
by nontrivial dependence on the radial coordinate of AdS and of the Sphere respectively. 
A kernel defining a one to one map between these states is
then constructed. 
\end{abstract}

\newpage
\noindent
\section{Introduction}
 Studies [1-8] of giant gravitons  in AdS Supergravity (and dual N=4 SYM theory) have lead to a
simple (matrix model) picture for 1/2 BPS states. In particular a free fermion model [8,9,10] of
harmonic oscillators was identified and shown to simulate fully the dynamics of 
1/2 BPS states and  their interactions. In  [10] (referred to as LLM) a  classical Ansatz for AdS 
(bubbling) configurations was constructed whose energy and flux were demonstrated to be
in a one to one correspondence with those of a general fermionic droplet configuration. Further,
relevant studies of this free fermion map have recently been carried out [11-33].
 
 It is clear that it would be desirable to extend the map to more general states and go beyond the simple case of free fermions. This would require an investigation (and solution) of more complex two (or multi) matrix models, a formidable task. 
In the present work we present a step in this direction. We will attempt to extend the correspondence from the fermionic family of states (representing a single diagonal matrix) to a more general set associated with states of a  two matrix quantum mechanics. As was already seen in [8]  which concerned itself with the case of 1/2 BPS states one can start with a system of two matrices, or 
a complex matrix and  perform a truncation to a single hermitean matrix (in the manner analogous to a similar phenomena in the quantum Hall effect).
The reduction was explained in [8] to be the Hilbert space equivalent of a holomorphic projection where the set of observables are given by traces of the complex matrix Z only. The introduction of mixed traces, involving the second 
(conjugate) matrix immediately leads to a nontrivial dynamical problem whose eigenstates were never constructed. 

 We will first adress this problem of constructing invariant eigenstates of the two matrix 
quantum system. For this we develop in some detail a hybrid formalism, treating one of the  matrices fully in the standard collective field theory manner, while the other is treated in the coherent state representation. This second matrix behaves then as an 'impurity'. The corresponding collective field theory of combined, mixed traces
is then worked out and is shown to lead to a sequence of eigenvalue equations. These equations are seen to generalize an
eigenvalue equation first found in \cite{MarchesiniYQ}, and first solved
for its eigenstates in \cite{Halpern:1981fj},
describing angular degrees of fredom of the single matrix model. 
The sequence of eigenvalue equations can be solved for the present case of the oscillator potential. It provides a two parameter set of energy (dilatation operator) eigenvalues and a corresponding $2$ dimensional space of eigenfunctions. 

 The central issue then becomes that of providing a correspondence between the eigenstates of the
matrix model and  states and eigenvalues in  Supergravity. Here we work in a linearized approximation specifying a class of fluctuations with matching quantum numbers. The wavefunctions,
in the AdS x S background are  nontrivial, being given by hypergeometric functions or corresponding special functions. Nevertheless, we describe a $1-1$ map between a two dimensional subset and the two dimensional set of wavefunctions given by the matrix model. This map involves a transformation introduced originaly in the context of the 2d Black hole and the corresponding matrix model \cite{JY}. This transform, appropriately interpreted, then provides a one to one map
between the gravity and matrix model wavefunctions. We emphasize that being one to one this map is different from the well known holographic projection. It is expected that further studies of the map will be of relevance for reconstructing AdS quantum mechanics.

 The outline of the paper is as follows. In Sect.2 we give a review of the simple matrix model and 
of the fermion map. In
Sect.3 we adress the two matrix problem describing its collective field formulation. We derive a sequence of eigenvalue equations and solve for eigenvalues and eigenfunctions.
In Sect.4 we consider the wavefunctions of the AdSxS SUGRA and specify the transform to the matrix model eigenstates. 
Several Appendices contain further details.

\section{Review}
We begin by reviewing and clarifying the existing map between the $\frac{1}{2}$ BPS SUGRA configurations and the states of the harmonic oscillator matrix model.

The matrix model degrees of freedom originate from a reduction of $N=4$ Super Yang-Mills theory on $R\times S^{3}$. The hamiltonian is therefore the dilatation operator and the Higgs fields become quantum mechanical matrix coordinates $\Phi_{a}\left(t\right)$, $a=1\ldots 6$.  For the study performed in the
present paper one can concentrate on the dynamics of two matrices  

\begin{equation*}
S=\frac{1}{2g_{YM}^{2}} \int\, dt \tr\left(\dot{\Phi}_{1}^{2} +\dot{\Phi}_{2}^{2}- \Phi_{1}^{2}
-\Phi_{2}^{2}-\frac{1}{2}\left[\Phi_{1},\Phi_{2} \right]^{2}\right).
\end{equation*}
The commutator interaction did not play a role in the $\frac{1}{2}$ BPS correspondence and in what follows
we will mainly concern ourselves with the simple quadratic harmonic oscillator model of two matrices 
\begin{equation*}
H=\frac{1}{2}\tr\left(P_{1}^{2}+ P_{2}^{2}+\Phi_{1}^{2} +\Phi_{2}^{2} \right)
\end{equation*}
The symmetries of this reduced theory are given by the $U\left(1\right)$ charge
\begin{equation*}
J=\tr\left(P_{1}\Phi_{2}- P_{2}\Phi_{1} \right)
\end{equation*}
and an $SL\left(2,R\right)$ symmetry algebra (allternatively SU(2)).
One has the complex matrices
\begin{equation*}
\begin{aligned}
Z&=\frac{1}{\sqrt{2}}\left(\Phi_{1}+i\Phi_{2} \right)\\
Z^{\dag}&=\frac{1}{\sqrt{2}}\left(\Phi_{1}-i\Phi_{2} \right)
\end{aligned}
\end{equation*}
and the conjugates
\begin{equation*}
\begin{aligned}
\Pi&=\frac{1}{\sqrt{2}} \left(P_{1}+iP_{2} \right)=-i\frac{\partial}{\partial Z^{\dag}}\\
\Pi^{\dag}&=\frac{1}{\sqrt{2}} \left(P_{1}-iP_{2} \right)=-i\frac{\partial}{\partial Z}.
\end{aligned}
\end{equation*}
Restriction to $\frac{1}{2}$ BPS configurations corresponds in the matrix model to considering a subset
of correlators involving only the chiral primary operators of the general form
\begin{equation*}
\tr Z^{k_{1}}\tr Z^{k_{2}}\cdots \tr Z^{k_{n}}.
\end{equation*}
For the corresponding reduction in Hilbert space one proceeds as follows(see [8,9]).
It is useful to introduce the operators
\begin{equation*}
\begin{aligned}
A&=\frac{1}{2}\left(Z+i\Pi \right)\\
\end{aligned}
\end{equation*}
and 
\begin{equation*}
\begin{aligned}
B&=\frac{1}{2}\left(Z-i\Pi \right)\\
\end{aligned}
\end{equation*}

In terms of these, the Hamiltonian and the $U\left(1\right)$ charge read
\begin{equation*}
\begin{aligned}
H&=\tr\left(A^{\dag}A+B^{\dag}B \right)\\
J&=\tr\left(A^{\dag}A-B^{\dag}B \right)
\end{aligned}
\end{equation*}
One now has a  sequence of eigenstates given by
\begin{equation*}
\begin{aligned}
&\tr\left(\left(A^{\dag}\right)^{n}\right)\left|0\right\rangle \,\, E=J=n\\
&\tr\left(\left(B^{\dag}\right)^{n}\right)\left|0\right\rangle \,\, E=-J=n\\
&\tr\left(\left(A^{\dag}\right)^{n}\right) \tr\left(\left(B^{\dag}\right)^{m}\right)\left|0\right\rangle \,\,
E=n+m,\, J=n-m
\end{aligned}
\end{equation*}
Restriction to $\frac{1}{2}$ BPS configurations corresponds in the matrix model Hilbert space to
a reduction to a  subsector given by A oscillators. 
It is useful to diagonalize $A,A^{\dag}$ by using the unitary symmetry
\begin{equation*}
\begin{aligned}
&A_{ij}= \lambda_{i}\delta_{ij}\\
&A_{ij}^{\dag}=\lambda_{i}^{\dag} \delta_{ij}
\end{aligned}
\end{equation*}
 The measure in these variables shows that we can treat the $\lambda_{i}$'s as fermionic variables. The Hamiltonian for these fermionic oscillators is
\begin{equation*}
H=\sum_{i}\lambda^{\dag}_{i}\lambda_{i}.
\end{equation*}
The fermionic wavefunctions are
\begin{equation*}
\psi_{F}\left(\lambda_{1},\lambda_{2},\ldots,\lambda_{n} \right)= e^{-\sum_{i}\bar{\lambda}_{i}\lambda_{i}}\det \left(
\begin{array}{cccc}
\lambda^{l_{1}}_{1} & \lambda^{l_{2}}_{1} & \cdots & \lambda^{l_{N}}_{1} \\
\lambda^{l_{1}}_{2} & \lambda^{l_{2}}_{2} & \cdots & \lambda^{l_{N}}_{2} \\
\vdots & \vdots & \vdots & \vdots\\
\lambda^{l_{1}}_{N} & \lambda^{l_{2}}_{N} & \cdots & \lambda^{l_{N}}_{N}
\end{array}
\right).
\end{equation*}
After dividing the wavefunction by the Vandermonde determinant we have that
\begin{equation*}
\begin{aligned}
\psi_{B;l_{1},l_{2},\ldots,l_{N}}\left(\lambda_{1},\lambda_{2},\ldots,\lambda_{n} \right)&= e^{-\sum_{i}\bar{\lambda}_{i}\lambda_{i}} \chi_{l_{1},l_{2},\ldots,l_{N}}\left(\lambda_{1},\lambda_{2},\ldots,\lambda_{N}\right)
\end{aligned}
\end{equation*}
Where $\chi_{l_{1},l_{2},\ldots,l_{N}}$ denotes the character of a representation of $SU\left(N\right)$ that corresponds to a Young tableaux with $l_{1}$ boxes in the first row, $l_{2}$ boxes in the second one etc.  
 Of special interest is the sequence of states  corresponding  to  representations that contains 1 row of $l$ boxes
\begin{equation*}
\psi_{B;l_{1},l_{2},\ldots,l_{N}}\left(\lambda_{1},\lambda_{2},\ldots,\lambda_{n} \right)=e^{-\sum_{i}\bar{\lambda}_{i}\lambda_{i}} \chi_{l,0,\ldots,0}= e^{-\sum_{i}\bar{\lambda}_{i}\lambda_{i}} \chi_{l,0,\ldots,0}\left(\lambda_{1},\lambda_{2},\ldots,\lambda_{N}\right)
\end{equation*}
and another sequence that corresponds to a representation that contains 1 column of $l$ boxes
\begin{equation*}
\psi_{B;l_{1},l_{2},\ldots,l_{N}}\left(\lambda_{1},\lambda_{2},\ldots,\lambda_{n} \right)=e^{-\sum_{i}\bar{\lambda}_{i}\lambda_{i}} \chi_{l,0,\ldots,0}= e^{-\sum_{i}\bar{\lambda}_{i}\lambda_{i}} \chi_{1,1,\ldots,1,0,\ldots,0}\left(\lambda_{1},\lambda_{2},\ldots,\lambda_{N}\right).
\end{equation*}
In the fermionic picture \cite{AJnp} the first set of states represents particles and the second holes. These were explained 
in [8,6,9] to corresponds to a giant gravitons in $AdS$ and to a giant gravitons on the sphere respectively. In terms of the moments
\begin{equation*}
\phi_{i}=\sum_{j=0}^{N}\lambda_{j}^{i}
\end{equation*}
one obtains Schur polynomials representing these states
\begin{equation*}
\begin{aligned}
\chi_{l,0,\ldots,0}\left(\lambda_{1},\lambda_{2},\ldots,\lambda_{N}\right)&= P_{l}\left(\phi_{1},\phi_{2},\ldots,\phi_{N}\right)\\
\chi_{1,1,\ldots,1,0,\ldots,0}\left(\lambda_{1},\lambda_{2},\ldots,\lambda_{N}\right)&=\left(-\right)^{l}P_{l}\left(-\phi_{1},-\phi_{2},\ldots,-\phi_{N}\right)
\end{aligned}
\end{equation*}
They are exact eigenstates of a cubic collective field theory representing the bosonized
version of 1d fermions. In terms of a two dimensional  density field $\rho\left(x,y,t\right)$ the hamiltonian is simply
\begin{equation*}
H=\frac{1}{2}\int dx\int dy \left(x^{2}+y^{2}\right)\rho\left(x,y,t\right).
\end{equation*}
Together with the non-trivial symplectic form
\begin{equation*}
L_{0}=2\pi \int dx\int dy \rho\left(x\right)G\left(x-x^{\prime}\right) \dot{\tilde{\rho}}\left(x^{\prime}\right)
\end{equation*}
one has a topological $2+1$ dimensional scalar field theory \cite{IKS} which can be reduced to a $1+1$ dimensional collective field theory describing the dynamics of the boundary (of the droplet) $y_{\pm}\left(x,t\right)$ by
\begin{equation*}
L=\frac{1}{2\pi}\int dt\int dx \left[y_{+}\partial^{-1}_{x}
\dot{y}_{+}- y_{-}\partial^{-1}_{x} \dot{y}_{-}
-\left(\left(y_{+}^{3}-y_{-}^{3}\right)
+x^{2}\left(y_{+}-y_{-}\right)\right)\right]
\end{equation*}
One can parametrize the boundary in terms of radial coordinates, in which the Lagrangian becomes quadratic. This is a simple manifestation of the integrability of this theory. This goes as follows:

Consider a closed curve $\vec{r}\left(s,t\right)$ in $R^{2}$ with
parameter $s$ which in our case describes the boundary of the fermi
sea in the phase space. In general the equation of motion can be
written in the form
\begin{eqnarray*}
\partial_{t}\vec{r}\times \partial_{s}\vec{r}=\partial_{s}A\left(\vec{r}\right)
\end{eqnarray*}
with $A\left(\vec{r}\right)$ defining the model that we are
studying. For the case of free fermions in an oscillator potential
one has 
\begin{eqnarray*}
A\left(\vec{r}\right)= \frac{1}{2} \vec{r}^{2}.
\end{eqnarray*}
If we
parametrize the curve as
\begin{eqnarray*}
\vec{r}\left(x,t \right)=x\, \hat{x}+ y_{\pm}\left(x,t \right)\,
\hat{y}
\end{eqnarray*}
 we recover 
\begin{eqnarray*}
\partial_{t}\vec{r}\times \partial_{x}\vec{r}&=&\partial_{t}y_{\pm}\\
\partial_{t}y_{\pm} &=& -\frac{1}{2} \partial_{x}\left(y_{\pm}^{2}+x^{2} \right)
\end{eqnarray*}
If instead one uses polar coordinates to parametrize the 
boundary 

\begin{eqnarray*}
\vec{r}\left(\phi,t \right)=\rho\left(\phi,t
\right)\cos\left(\phi\right)\, \hat{x}+ \rho\left(\phi,t
\right)\sin\left(\phi\right)\, \hat{y}
\end{eqnarray*}
and in this case we have
\begin{eqnarray*}
\partial_{t}\vec{r}\times \partial_{\phi}\vec{r}&=&\frac{1}{2}
\partial_{t} \rho^{2}\left(\phi,t\right)\\
\partial_{t}\rho^{2}=\partial_{\phi}\rho^{2}.
\end{eqnarray*}
It is instructive  to derive the above linear equation of
motion from the non-linear one by using the field dependent
coordinate tranasformation.It is simply given by
\begin{eqnarray*}
x&=&\rho\left[\phi\left(x,t\right),t \right]
\cos\left(\phi\left(x,t\right) \right)\\
y_{+}&=& \rho\left[\phi\left(x,t\right),t \right]
\sin\left(\phi\left(x,t\right) \right)
\end{eqnarray*}

These are then the action-angle coordinates for the dynamics of the
boundary.

 In their work Lin, Lunin and Maldacena \cite{Lin:2004nb} have identified a nonlinear Ansatz for $10$ d Sugra 
which exactly reduces to 
the above, bosonic hamiltonian of 1d fermions . To summarize the main 
features of the Ansatz, one has first the 10 dimensional metric

\begin{equation*}
\begin{aligned}
ds^2 &= - h^{-2}
(dt + V_i dx^i)^2 + h^2 (dy^2 + dx^idx^i) + y e^{G
 } d\Omega_3^2 + y e^{ - G} d \tilde \Omega_3^2
\\
h^{-2} &= 2 y \cosh G ,
\\
 y \partial_y V_i &= \epsilon_{ij} \partial_j z,\qquad
 y (\partial_i V_j-\partial_j V_i) = \epsilon_{ij} \partial_y z
 \\
 z &=\frac{1}{2} \tanh G
 \\
\end{aligned}
\end{equation*}

and a corresponding Ansatz for the the gauge fields.

The only unknown function z is shown to obey the Laplace equation:

\begin{equation*}
\partial_i \partial_i z + y \partial_y ( { \partial_y z \over y} ) =0
\end{equation*}
which is solved as a boundary value problem :
\begin{equation*}
\begin{aligned}
z(x_1,x_2,y)&=\frac{y^2}{\pi}\int_{\cal D} \frac{z(x_1',x'_2,0)
dx_1'dx_2'}{[({\bf x}-{\bf x}')^2+y^2]^2}
\\
V_i(x_1,x_2,y) &= { \epsilon_{ij} \over \pi} \int_{\cal D}
\frac{z(x_1',x'_2,0) (x_j - x'_j) dx_1'dx_2'}{[({\bf x}-{\bf
x}')^2+y^2]^2}
\end{aligned}
\end{equation*}
Remarkably,the flux and the energy of this general configuration were shown by LLM to take the form of the bosonized free fermion droplet
\begin{equation*}
N=\frac{1}{4\pi^{2}l_{P}^{2}}\int dx_{1} \int dx_{2} \left(u\left(t,x_{1},x_{2}\right)+\frac{1}{2} \right)
\end{equation*}
\begin{equation*}
\begin{aligned}
\Delta=&\frac{1}{4\pi \hbar^{2}}\int dx_{1}\int dx_{2} \left( x_{1}^{2}+ x_{2}^{2}\right) \left(u\left(t,x_{1},x_{2}\right)+\frac{1}{2} \right)\\
-&\frac{1}{8\pi^{2} \hbar^{2}} \left( \int dx_{1}\int dx_{2} \left( x_{1}^{2}+ x_{2}^{2}\right) \left(u\left(t,x_{1},x_{2}\right)+\frac{1}{2} \right) \right)^{2}
\end{aligned}
\end{equation*}
It should be stressed that even though these expressions look two dimensional, effectively this is still only a 
$1$ dimensional correspondence (it is described explicitely by the 1+1 dimensional bosonic scalar field theory).  In addition to the formulas for the flux and the energy one also needs the symplectic form (which should coincide
with the symplectic form established by Iso,Karabali and Sakita \cite{IKS}) for the 2d fermion droplet. Another, 
simple way to see the one dimensionality is by an analysis of linearized fluctuations (we give this in Appendix A). One has
\begin{equation*}
S=\sum_{n>0}\frac{1}{2}\int dt \left[\frac{1}{n^2}\dot{p}_{n}^{2} +
\dot{q}^{2}_{n}-n^{2}q_{n}^{2}-p_{n}^{2} \right]
\end{equation*}

in agreement with the well known quadratic action for chiral primaries in AdS:

\begin{equation*}
S=\sum_{n}\frac{8R^{8}_{AdS} n \left(n-1
\right)}{\left(n+1\right)^{2}}
\int_{AdS^{5}}dx^{5}\sqrt{g_{AdS^{5}}}\left[\sigma^{-n}\Box
 \sigma^{+n} - n\left(n-4 \right)\sigma^{-n} \sigma^{+n} \right]
\end{equation*}
It is supersymmetry which requires that
$\left(\partial_{t}-\partial_{\phi}\right)\sigma=0$ which for the
$0+1$ dimensional variables means that the "angular momentum" is
equal to the energy. Choosing an opposite chirality for the fermions
we would have had the condition
$\left(\partial_{t}+\partial_{\phi}\right)\sigma=0$ which would flip
the sign in the relation between energy and "angular momentum".
\noindent
\section{Matrix Model Eigenproblem}

We have seen in the discussion that the treatment of 1/2 BPS states corresponds
to a reduction, namely to one matrix quantum mechanics given by the canonical
set $A$ and $A^{\dag}$. It is our interest to extend this correspondence to a larger set of states. 
In the matrix model they will be states involving the two matrices ($A$ and $B$) of a two matrix model.
This can be stated as a two matrix problem, with two  hermitian matrices M and N in a quadratic potential,
i.e., with Hamiltonian

\bea
H
\equiv
&-&{1\over 2}
Tr ({\partial \over \partial M} {\partial \over \partial M}) +
{1\over 2} Tr (M^2) -{1\over 2}
Tr ({\partial \over \partial N} {\partial \over \partial N}) +
{1\over 2} Tr (N^2)
\eea

\noindent
Using creation-annhilation operators for the matrix $N_{ij}$ in a coherent basis, the Hamiltonian
takes the form considered in this article:

\be
\hat{H}\equiv - {1\over 2}
Tr ({\partial \over \partial M} {\partial \over \partial M}) +
{1\over 2} Tr (M^2) + Tr (B {\partial \over \partial B})
\ee

\noindent
We consider the action of this hamiltonian on functionals of invariant variables (loops)

$$
\Phi \Big[ \psi(k,s=0,1,2,...) \Big],
$$

\noindent
where the $\psi(k,s=0,1,2,...)$ are states with $s$ "$B$ impurities":

\bea
\psi(k,0)  &= & Tr(e^{ikM})   \nonumber\\
\psi(k,1)  &=  & Tr(B e^{ikM})\nonumber  \\
\psi(k,2)  &= & \int_0^k dk' Tr(B e^{ik'M} B e^{i(k-k')M})\\
           & ... \nonumber
\eea

\noindent
In terms of the eigenvalues $\lambda_i$ and the angular variables $V$ of the matrix $M=V\Lambda V^{+}$, we have

\bea \label{LoopDefEig}
\psi(k,0) &= &\Sigma_{i} e^{ik\lambda_i} \nonumber\\
\psi(k,1) &=& \Sigma_{i}  (V^{+}B V)_{ii} e^{ik\lambda_i} \\
\psi(k,2) &=& -2i \Sigma_{i,j}(V^{+}B V)_{ij} (V^{+}B V)_{ji} {e^{ik\lambda_j}\over (\lambda_j-\lambda_i)}\nonumber\\
& ...& \nonumber
\eea

\noindent
Using the chain rule, we obtain for the matrix $M$ kinetic energy operator on the wave functional:

\beas
&-&{1 \over 2 }
Tr ({\partial \over \partial M} {\partial \over \partial M})
=-{1 \over 2 }\Sigma_{s} \int dk \quad
Tr({\partial^2 \psi(k,s) \over \partial M \partial M})
{\partial \over \partial \psi(k,s)} \\
&-&{1 \over 2 } \Sigma_{s,s'}\int dk \int dk'\quad
Tr({\partial\psi(k,s) \over \partial M}{\partial\psi(k',s') \over \partial M})
{\partial^2 \over \partial \psi(k,s) \partial \psi(k',s')}
\eeas

\noindent
As it is traditional \cite{JevickiMB}, we introduce the notation:

\bea \label{LapK}
-{1 \over 2 }Tr({\partial \over \partial M} {\partial \over \partial M})=
&-&{1 \over 2 } \Sigma_{s} \int dk \quad \omega (k,s) {\partial \over \partial \psi(k,s)}  \\
&-&{1 \over 2 }\Sigma_{s,s'}\int dk \int dk'\quad
\Omega (k,s:k',s')
{\partial^2 \over \partial \psi(k,s) \partial \psi(k',s')}
\nonumber
\eea

\noindent
$\omega (k,s)$ splits the loop $\psi(k,s)$ and $\Omega (k,s:k',s')$ joins the two loops $\psi(k,s)$ and $\psi(k',s')$.

We will find it useful to introduce a density description, or $x$ representation:

$$
\psi(x,s) = \int {dk \over 2 \pi} e^{-ikx} \psi(k,s),  \qquad \psi(k,s) = \int {dx} e^{ikx} \psi(x,s).
$$

\noindent
Any function of $k$ (or $x$) transforms accordingly. Namely:

\beas
\omega(x,s) &=& \int {dk \over 2 \pi} e^{-ikx} \omega(k,s) \\
\Omega(x,s;y,s') &= &\int {dk \over 2 \pi} \int {dk' \over 2 \pi} e^{-ikx} e^{-ik'y} \Omega(k,s;k',s')
\eeas

\noindent
For conjugates, we have

$$
{\partial \over \partial \psi(x,s)}= \int dk e^{ikx} {\partial \over \partial \psi(k,s)}; \quad
{\partial \over \partial \psi(k,s)} = \int {dx \over 2 \pi} e^{-ikx} {\partial \over \partial \psi(x,s)}
$$

\noindent
In the density description, the kinetic operator then becomes:

\bea \label{LapX}
& & {-{1\over 2}} Tr({\partial \over \partial M} {\partial \over \partial M})
  = {-{1\over 2}}\Sigma_{s} \int dx \omega (x,s) {\partial \over \partial \psi(x,s)} \\
& &{-{1\over 2}}\Sigma_{s,s'} \int dx \int dy \Omega(x,s:y,s')
{\partial^2 \over \partial \psi(x,s) \partial \phi(y,s')} \nonumber
\eea

\subsection{Spectrum and fluctuations in the zero impurity sector}
\noindent
Consider first the analysis for the spectrum of the zero impurity problem. This sector corresponds 
to the Quantum Mechanics of a single hermitean matrix, and it has by now a standard
solution \cite{JevickiMB},\cite{DasKA},\cite{DemeterfiCW}, which is briefly reviewed in Appendix B. 
In this case, one has the standard cubic Hamiltonian 

\be \label{HEffZero}
H_{eff}^{0}= {{1 \over 2N^2}}
\int dx \partial_x \Pi(x) \psi(x,0) \partial_x \Pi(x) +
N^2 \Big(  \int dx {\pi^2 \over 6}\psi^3(x,0) +  \psi(x,0)({x^2 \over 2}- \mu) \Big)
\ee

\noindent
giving the well known Wigner distribution background in the limit as $N \to \infty$

$$
\pi \psi(x,0) \equiv \pi \phi_0 = \sqrt{2\mu - x^2} = \sqrt {2-x^2}.
$$

\noindent
For the small fluctuation  spectrum, one shifts the background

$$
\psi(x,0) = \phi_0 + {1\over \sqrt{\pi} N} {\partial_x \eta };
\qquad \partial_x \Pi(x) = - \sqrt{\pi} N P (x)
$$

\noindent
to find the quadratic operator

$$
H_{2}^{0}= {{1 \over 2}}
\int dx \pi \phi_{0}P^2(x) + {1 \over 2} \int dx \pi\phi_0 ({\partial_x \eta })^2
$$

\noindent
The way to diagonalize is by now well known: one changes to the classical "time of flight" $q$

$$
{dx \over dq} = \pi \phi_0 ; \qquad x(q)= - \sqrt{2} \cos(q) ; \qquad \pi \phi_0 =  \sqrt{2} \sin (q);
\quad0 \le q \le \pi
$$

\noindent
One obtains the equation for a $2d$ massless boson:

\be\label{Quadratic}
H_{2}^{0}= {{1 \over 2}}
\int dq P^2(q) + {1 \over 2} \int dq ({\partial_q \eta })^2
\ee

In addition one needs to impose Dirichelet boundary conditions at the classical turning points, for a consistent
time evolution of the constraint (\ref{Const}). Therefore the spectrum in the zero impurity sector is

\be \label{SpecZero}
                  w_j= j  \quad ; \qquad     \phi_j = \sin(jq)
\ee

The following comment is in order: the harmonic oscillator potential is special, in that the effective hamiltonian
(\ref{HEffZero}) can be equivalently written as (for discussions on the relationship between the two re-writings in the
context of supersymmetric or stochastic stabilizations, see for instance  \cite{JevickiYK}, \cite{RodriguesBY}, \cite{vanTonderVC},\cite{Alastair})

$$
H_{eff}^{0}= {{1 \over 2N^2}}
\int dx \partial_x \Pi(x) \psi(x,0) \partial_x \Pi(x) +
{N^2 \over 2} \int dx \psi(x,0)  \big( \int dy {\psi(y,0) \over x-y} - x \big)^2 .
$$

It is then seen that the Wigner distribution background also safisfies the well known BIPZ \cite{BrezinSV} equation

\be \label{BIPZ}
\int dz {\phi_0(z) \over (x-z)}=x
\ee

\noindent
Shifting about the background as above, we obtain for the quadratic hamiltonian

$$
H_{2}^{0}= {{1 \over 2}}
\int dx \pi \phi_{0}P^2(x) + {1 \over 2} \int dx \pi\phi_0
\Big( \int {dy \over \pi} {\partial_y \eta(y) \over x-y} \Big)^2
$$

\noindent
This non-local hamiltonian can be easily shown to be equivalent to (\ref{Quadratic}). Let us examine this
in slightly more detail: by changing to the  classical time of flight $q$, we obtain

$$
H_{2}^{0}= {{1 \over 2}}
\int dq P^2(q) + {1 \over 2} \int dq
\Big( \partial_q \int {dq' \over \pi} {\pi \phi_0 (q') \eta(q') \over x(q)-x(q')} \Big)^2
$$

\noindent
The above non local integral operator plays a prominent role in what follows and is discussed in Appendix C.
Let us denote it by

$$
\partial_q \int {dq' \over \pi} {\pi \phi_0 (q') f(q') \over x(q)-x(q')} \equiv - i |\partial_q| f(q)
$$

\noindent
and by abuse of language (it does not satisfy a Leibnitz rule) refer to it as the "absolute derivative", for ease 
of notation. We note that
$(- i |\partial_q|)^2=\partial_q^2$ and that the appropriate eigenfunctions of this operator
are $\phi_n = \sin(nq)$ with eigenvalue $n$ as shown in Appendix C. Therefore the eigenfunctions (\ref{SpecZero})
are also the solutions of

$$
           (i\partial_t  +  i |\partial_q|) \phi(q)=0
$$

\subsection{Quadratic hamiltonian for states with impurities}

\noindent
We return now to the (pre-hermitean) kinetic energy operator (\ref{LapX}) (or (\ref{LapK})). We note that

$$
<\psi(x,s)> = <\psi(k,s)> = 0 \quad ; \qquad s=1,2,3,...
$$

\noindent
This observation implies that for the multi-impurity spectrum it is sufficient to consider
the zero impurity sector Jacobian already discussed \cite{AvJe}, i.e.,

\beas
& & {\partial \over \partial \psi(x,0)}
\to J^{1\over 2} {\partial \over \partial \psi(x,0)}J^{-{1\over 2}} =
{\partial \over \partial \psi(x,0)} - {1\over 2}{\partial \over \partial \psi(x,0)} \ln J \\
& &{\partial \over \partial \psi(x,s)} \to {\partial \over \partial \psi(x,s)}, \quad s=1,2,3,...
\eeas

\noindent
where, to leading order in $N$

\be \label{Jac}
\partial_x {\partial \ln J \over \partial \psi(x,0)} = \partial_x \int dy \Omega^{-1}(x,0;y,0) \omega(y,0) =
2 \int dy {\psi(y,0) \over (x-y)}
\ee

\noindent
Let us now identify the terms in (\ref{LapX}) which determine the quadratic operator in the multi-impurity sector.

\noindent
We look for terms of the form $\psi(x,s) \partial / \partial \psi(x,s), s>0$ when $\psi(x,0) \to \phi_0(x)$.
Contributions of this form contained in the first term of (\ref{LapX}) result from 
splittings of the loop $\psi(x,s)$ into
a zero impurity loop and another with $s$ impurities. We will denote this amplitude by $\bar{\omega}(x,s)$.

\noindent
Contributions contained in the second term of (\ref{LapX}) are obtained as a 
result of the similarity transformation described above, when we replace $\partial / \partial \psi(x,0) \to  - (1 / 2) \partial / \partial \psi(x,0) \ln J$. 
We therefore
obtain:

\be \label{HTwoMulti}
H_2^{s} = {-{1\over 2}} \int dx \bar{\omega} (x,s) {\partial \over \partial \psi(x,s)}
+ {1\over 2} \int dx \int dy \Omega(x,0:y,s)
          {\partial \ln J \over \partial \psi(x,0)} {\partial \over \partial \psi(y,s)}
\ee

In a problem involving joining and splitting of loop states, the issue of closure of loop space is an important
one. The first term in (\ref{HTwoMulti}) always closes. This is because

$$
   \bar{\omega} (k,s) = -2 \int_0^k dk' k'   \psi(k',s) \psi(k-k',0)
$$

\noindent
This result is a straightforward application of a result established in \cite{deMelloKochNQ}. 
In the $x$ representation,

\beas
& &\bar{\omega}(z,s) = \int {dk \over 2 \pi} e^{-ikz} \bar{\omega}(k,s) \\
& &= - 2 \Big[\psi (z,s) \int dx {\phi_0(x) \over (x-z)^2} - \phi_0(z) \int dx {\psi(x,s) \over (x-z)^2}
+ \int dx \phi_0(x) \partial_z \big( {\psi(z,s) \over (z-x)} \big)\Big]
\eeas

\noindent
Substituting this expression into (\ref{HTwoMulti}) we obtain

\begin{eqnarray} \label{HTwoOme}
H_2^{s}&=& \int dx \int dz {\phi_0(z)\psi(x,s)- \psi(z,s) \phi_0(x) \over (x-z)^2} 
{\partial \over \partial \psi(x,s)} \nonumber \\
&- &\int dx \int dz  {\phi_0(z) \psi(x,s) \over (x-z)} \partial_x {\partial \over \partial \psi(x,s)}  \\
&+& {1\over 2} \int dx \int dy \Omega(x,0:y,s)
          {\partial \ln J \over \partial \psi(x,0)} {\partial \over \partial \psi(y,s)} \nonumber
\end{eqnarray}

\noindent
In general, for an arbitrary potential, the last term in (\ref{HTwoOme}) involving $\Omega(x,0:y,s)$ will not close.
We will argue in the following that for the harmonic oscillator potential this term closes, by considering
explicitely $s=1,2,3$ and then arguing for the general case.

\subsection{The one impurity sector}

\noindent
It is straighforward to show that in this case

$$
\Omega(k,0:k',1) = - k k' \psi (k+k',1)
$$

\noindent
from which it follows

$$
\Omega(x,0;y,1) = \partial_x \partial_y (\psi (x,1) \delta (x-y)).
$$

\noindent
The term involving $\Omega(x,0:y,1)$ in (\ref{HTwoOme}) becomes

\beas
& &{1\over 2} \int dx \int dy \Omega(x,0:y,1)
{\partial \ln J \over \partial \psi(x,0)} {\partial \over \partial \psi(y,1)} \\
& &= {1\over 2} \int dx \partial_x {\partial \ln J \over \partial \psi(x,0)} \psi (x,1) \partial_x
{\partial \over \partial \psi(x,1)} \\
& &=\int dx \int dz {\phi_0(z) \over (x-z)} \psi (x,1) \partial_x {\partial \over \partial \psi(x,1)},
\eeas

\noindent
where we have used (\ref{Jac}). We observe that this term cancels exactly a similar term in (\ref{HTwoOme}) , and we obtain
the final form for the quadratic hamiltonian in the $1$ impurity sector:

\be
H_2^{s=1} = \int dx \int dz {\phi_0(z)\psi(x,1)- \psi(z,1) \phi_0(x) \over (x-z)^2} {\partial \over \partial \psi(x,1)}
\ee

\noindent
The rescaling (\ref{Rescaling}) leaves the above hamiltonian invariant, or equivalently the above hamiltonian is of
order $1$ ($N^0$) in $N$, as was the case in the zero impurity sector. Writing the operator as

$$
\int dx \int dy \psi(x,1) K(x,y) {\partial \over \partial \psi(y,1)},
$$

\noindent
we obtain
$$
\int dy K (x,y) {\partial \over \partial \psi(y,1)} =
 \int dy { \phi_0(y) \over (x-y)^2}
\Big( {{\partial \over \partial \psi(x,1)}- {\partial \over \partial \psi(y,1)}} \Big)
$$

\noindent
Acting on a wave functional

$$
          \Phi = \int dz f(z) \psi(z,1) ,
$$

\noindent
we obtain the Marchesini-Onofri kernel \cite{MarchesiniYQ},\cite{Halpern:1981fj},\cite{GK},\cite{Mls}

$$
\int dy { \phi_0(y) \over (x-y)^2}
\Big( f(x)- f(y) \Big) =  \Big( -{d \over dx} \int dy { \phi_0(y) \over (x-y)} \Big) f(x) +
{d \over dx} \int dy { \phi_0(y) f(y) \over (x-y)}
$$

\noindent
Using (\ref{BIPZ}), the first term yields $-f(x)$, and by changing to the time of flight coordinates the kernel can be
written as

$$
- f(q)  - {i \over \pi \phi_0}  |\partial_q| (\pi \phi_0(q) f(q)),
$$

\noindent
or, for the spectrum equation

$$
 ( - 1 - i  |\partial_q|)  (\pi \phi_0(q) f(q)) = w  (\pi \phi_0(q) f(q)) .
$$

\noindent
As described in Appendix C the spectrum and eigenfunctions of this operator are

$$
                    w_n=n-1 \quad ; \qquad \phi_n^{s=1}= {\sin (nq) \over \sqrt{2} \sin(q)} \quad ; \quad n=1,2,...
$$

\noindent
For the the harmonic oscillator potential, these are the well known Tchebychev polynomials of the second kind.
Adding the contribution from the $Tr (B {\partial / \partial B}) $ term of the Hamiltonian we obtain

\be
w_n= n \quad ; \qquad \phi_n^{s=1}= {\sin (nq) \over \sqrt{2} \sin(q)} \quad ; \quad n=1,2,...
\ee

\subsection{The two impurities sector}

\noindent
For two impurities, we have

\beas
\Omega(k_0,0:k,2) &=& -2 k_0 \int dk'k' Tr(B e^{i(k-k')M} B e^{i(k'+k_0)M})\\
 &=&
-2 k_0 \Sigma_{i,j} (V^{+}B V)_{ij} (V^{+}B V)_{ji} \Big[
-i k { e^{i(k+k_0)\lambda_i}  \over (\lambda_i - \lambda_j)} + e^{i k_0 \lambda_i}
{e^{i k \lambda_i} - e^{i k \lambda_j} \over (\lambda_i - \lambda_j)^2} \Big]
\eeas

\noindent
and

\beas
\Omega(x,0:y,2)= &-2 i \partial_x \partial_y \Sigma_{i,j} (V^{+}B V)_{ij} (V^{+}B V)_{ji} \delta(x-y)
{\delta(y-\lambda_i) \over (\lambda_i - \lambda_j)} \\
&-2 i \partial_x  \Sigma_{i,j} (V^{+}B V)_{ij} (V^{+}B V)_{ji} \delta(x-\lambda_i)
{\delta(y-\lambda_i) - \delta(y-\lambda_j)\over (\lambda_i - \lambda_j)^2}
\eeas

\noindent
The $\Omega(x,0:y,2)$ term in (\ref{HTwoOme}) takes the form

\beas
& &{1\over 2} \int dx \int dy \Omega(x,0:y,2)
{\partial \ln J \over \partial \psi(x,0)} {\partial \over \partial \psi(y,2)} \\
&=& - i \int dx \int dy \Big[
\Sigma_{i,j} (V^{+}B V)_{ij} (V^{+}B V)_{ji} \delta(x-y) {\delta(y-\lambda_i) \over (\lambda_i - \lambda_j)}
\partial_x {\partial \ln J \over \partial \psi(x,0)}
\partial_y {\partial \over \partial \psi(y,2)} \\
&-&\Sigma_{i,j} (V^{+}B V)_{ij} (V^{+}B V)_{ji} \delta(x-\lambda_i)
{\delta(y-\lambda_i) - \delta(y-\lambda_j)\over (\lambda_i - \lambda_j)^2}
\partial_x {\partial \ln J \over \partial \psi(x,0)}
{\partial \over \partial \psi(y,2)}
\Big]\\
&=&-2i \int dx \Sigma_{i,j} (V^{+}B V)_{ij} (V^{+}B V)_{ji} {\delta(x-\lambda_i) \over (x - \lambda_j)}
\int dz {\phi_0(z) \over (x-z)} \partial_x {\partial \over \partial \psi(x,2)}\\
&+& 2i \int dy \Sigma_{i,j} (V^{+}B V)_{ij} (V^{+}B V)_{ji}
{\delta(y-\lambda_i) \over (\lambda_i - \lambda_j)^2}
\Big[ \int dz {\phi_0(z) \over (\lambda_i-z)} - \int dz {\phi_0(z) \over (\lambda_j-z)}\Big]
{\partial \over \partial \psi(y,2)}
\eeas

\noindent
For the harmonic oscillator potential, we can use the result (\ref{BIPZ}), so that

\bea \label{Interm}
&&{1\over 2} \int dx \int dy \Omega(x,0:y,2)
{\partial \ln J \over \partial \psi(x,0)} {\partial \over \partial \psi(y,2)} \nonumber\\
& &=-2i \int dx \Sigma_{i,j} (V^{+}B V)_{ij} (V^{+}B V)_{ji} {\delta(x-\lambda_i) \over (x - \lambda_j)}
\int dz {\phi_0(z) \over (x-z)} \partial_x {\partial \over \partial \psi(x,2)}\nonumber\\
&&+2i \int dy \Sigma_{i,j} (V^{+}B V)_{ij} (V^{+}B V)_{ji}
{\delta(y-\lambda_i) \over (y - \lambda_j)} {\partial \over \partial \psi(y,2)}
\eea

\noindent
But from (\ref{LoopDefEig}),

\beas
\psi(x,2)& = & \int {dk \over 2 \pi} e^{-ikx} \psi(k,2) \\
& = &  -2i \int {dk \over 2 \pi} e^{-ikx} \Sigma_{i,j}(V^{+}B V)_{ij} (V^{+}B V)_{ji} {e^{ik\lambda_j}\over (\lambda_j-\lambda_i)}\\
& = & -2i \Sigma_{i,j}(V^{+}B V)_{ij} (V^{+}B V)_{ji} {\delta(x-\lambda_j)\over (x-\lambda_i)}
\eeas

\noindent
This allows us to express (\ref{Interm}) entirely in terms of the density $\psi(x,2)$ as

\beas
&{1\over 2}& \int dx \int dy \Omega(x,0:y,2)
{\partial \ln J \over \partial \psi(x,0)} {\partial \over \partial \psi(y,2)} \\
&=& \int dx \int dz {\phi_0(z) \over (x-z)} \psi(x,2)  \partial_x {\partial \over \partial \psi(x,2)}
- \int dx \psi(x,2) {\partial \over \partial \psi(x,2)}
\eeas

\noindent
As was the case for the one impurity sector, the first term above cancels the similar term in (\ref{HTwoOme}), 
and we obtain
for the quadratic hamiltonian in the $2$ impurity sector:

\be
H_2^{s=2} = \int dx \int dz {\phi_0(z)\psi(x,2)- \psi(z,2) \phi_0(x) \over (x-z)^2} {\partial \over \partial \psi(x,2)}
- \int dx \psi(x,2) {\partial \over \partial \psi(x,2)}
\ee

\noindent
This is a shifted Marchesini-Onofri operator. It can be recast in the form:

$$
 ( - 2 - i  |\partial_q|)  (\pi \phi_0(q) f(q)) = w  (\pi \phi_0(q) f(q)) .
$$

\noindent
The spectrum and eigenfunctions of this operator are

$$
        w_n=n-2 \quad ; \qquad \phi_n^{s=2}= {\sin (nq) \over \sqrt{2} \sin(q)} \quad ; \quad n=1,2,...
$$

\noindent
Adding the contribution from the $Tr (B {\partial / \partial B}) $ term of the Hamiltonian we obtain

\be
w_n= n \quad ; \qquad \phi_n^{s=2}= {\sin (nq) \over \sqrt{2} \sin(q)} \quad ; \quad n=1,2,...
\ee

\noindent

\subsection{Multi-impurity spectrum}

The pattern that emerges from the above discussion is clear: for $s$ impurities and the harmonic oscillator
potential, one obtains a shifted Marchesini-Onofri operator with spectrum and eigenfunctions

$$
                    w_n=n-s \quad ; \qquad \phi_n^{s}= {\sin (nq) \over \sqrt{2}\sin(q)} \quad ; \quad n=1,2,...
$$

\noindent
When the contribution from the $Tr (B {\partial / \partial B}) $ is added, we have for the full hamiltonian

\be
w_n= n \quad ; \qquad \phi_n^{s}= {\sin (nq) \over\sqrt{2} \sin(q)} \quad ; \quad n=1,2,...
\ee

\noindent
To provide further evidence of this pattern, the $3$ impurity case is treated explicitly in Appendix D. We
also checked that by introducing multi local densities and then projecting to the $2$ and $3$ impurity
states discussed here, we obtain the spectrum described above.

\noindent
To summarize, as the U(1) charge operator $\hat J$
it is represented by

$$ \hat J = - {1\over 2} Tr ({\partial \over \partial M} {\partial \over \partial M}) +
{1\over 2} Tr (M^2) - Tr (B {\partial \over \partial B}),
$$ and consequently $j=n-2s$. Together with the energy eigenvalues $w=n$, these specify a two parameter
family of states and a two dimensional complete set of eigenfunctions.

\section{SUGRA Map}
In this section we would like to identify the states of Sugra fluctuations and establish a 
one to one map with the eigenstates of the matrix problem found in the previous section. 
With the two matrices we hope to explore the extra coordinate which will be related to the radial coordinate of AdS and S. Since the other angular coordinates are ignored, it is sufficient to
concentrate on the small fluctuation equations asociated with $AdS_3 \times S_3$ (the analysis for $AdS_5 \times S_5$ 
reaches
an identical conclusion). We have obtained in the matrix model solution a two parameter sequence of states
with the eigenvalues $J=j$ and $w= j+2n$. It is easy to find a corresponding sequence of states, which
have the same eigenvalues. Actually there are two sequences ,one with nontrivial functional 
dependence in the radial variable of $AdS$ and the other in $S$. This situation is familiar from giant gravitons.

It will be  clear that while the integer valued eigenvalues easily agree (between the matrix model and supergravity), the comparison of their
 wavefunctions is much less trivial and also much more interesting. In Sugra the wavefunctions are given as nontrivial special
functions, while in the solution of the matrix eigenvalue problem they take the form of
ordinary plane waves . The later obviously happens after the change from eigenvalue coordinate to the 
"time of flight" coordinate. We will establish a  relationship between the two pictures in
terms of a kernel describing a (canonical) change of variables.

\subsection{The LLM kernel}
It is useful first to work out the form of the kernel for the case of $1/2$ BPS states given by the LLM map. For this one has
to consider the  LLM construction and perform the small fluctuation analysis . We do this in Appendix A
where we also give the details of a transformation to the Lorentz-De Donder gauge. Furthermore,
from now on the time of flight $q$ will be denoted by $\tau$. 

Let us concentrate on the
fluctuations associated with the metric $g_{\tilde{\Omega}\tilde{\Omega}}$. In the gauge of LLM 
the perturbation $\delta g_{\tilde{\Omega}\tilde{\Omega}}$ reads

\begin{equation}
\begin{aligned}
\delta g_{\tilde{\Omega}\tilde{\Omega}}&=-2 \sinh\rho \sin\theta
\sqrt{\frac{1+2u_{AdS}}{1-2u_{AdS}}}
\frac{1}{\left(1+2u_{AdS}\right)^{2}}\tilde{u}
d\tilde{\Omega}_{3}^{2}\\
&= \sin^2\theta \frac{1}{2\pi} \int_{0}^{2\pi} d\tau
\frac{\left(1-a^{2} \right)^{2}}{\left[1+a^{2}-2a\cos\left(\tau-
\phi \right) \right]^{2}}\sum_{j}a_{j}e^{ij\tau}\\
a&=\frac{\cos\theta}{\cosh\rho}
\end{aligned}
\end{equation}

The relevant  gauge transformation can be written in 
integral form as

\begin{equation}
\begin{aligned}
\delta\theta&=- \frac{\sin\theta
\cos\theta}{\cosh^{2}\rho-\cos^{2}\theta}
\sum_{j}a_{j}e^{ij\phi}\\
&=-\tan\theta\frac{1}{2\pi}\frac{a^{2}}{1-a^{2}}
\int_{0}^{2\pi}d\tau \frac{1-a^{2}}{1+a^{2}-2a\cos\left(\tau -\phi
\right)} \sum_{j}a_{j}e^{ij\tau}.
\end{aligned}
\end{equation}

Performing the  gauge transformation we have 
\begin{equation}
\delta g_{\tilde{\Omega}\tilde{\Omega}}= \sin^{2}\theta
\frac{1}{2\pi} \int_{0}^{2\pi}d\tau \frac{1-4a^{2}
-a^{4}+4a^{3}\cos\left(\tau-\phi \right)}{\left[1+a^{2}- 2a
\cos\left(\tau-\phi \right) \right]^{2}} \sum_{j}a_{j}e^{ij\tau}
\end{equation}
In this form we see the relation 
\begin{equation}
2\left|j\right|\sigma_{j}\left(t,\rho,\phi,\theta \right)=
\frac{1}{2\pi} \int_{0}^{2\pi}d\tau \frac{1-4a^{2}
-a^{4}+4a^{3}\cos\left(\tau-\phi \right)}{\left[1+a^{2}- 2a
\cos\left(\tau-\phi \right) \right]^{2}} \sum_{j}a_{j}e^{ij\tau}
\end{equation}

After performing the field dependent gauge transformation in order
to recognize the primary field coming from the metric and the three
form one has the relation

\begin{equation}
\begin{aligned}
\left|j\right|\sigma_{j}\left(t,\rho,\phi,\theta\right)=
\frac{1}{2\pi }e^{ijt} \int_{0}^{2\pi} d\tau K^{LLM}\left(\rho,
\phi,\theta|\tau \right) e^{ij\tau}
\end{aligned}
\end{equation}
where the kernel is given by
\begin{equation}
\begin{aligned}
&K^{LLM}\left(\rho, \phi,\theta|\tau \right)=\frac{1-4a^{2}
-a^{4}+4a^{3}\cos\left(\tau-\phi \right)}{\left[1+a^{2}- 2a
\cos\left(\tau-\phi \right) \right]^{2}}\\
&a=\frac{\cos\theta}{\cosh\rho}
\end{aligned}
\end{equation}
At this point we notice that $a<1$ at points where the measure
of $AdS_{3}\times S^{3}$ is non-zero. At this places we 
introduce a cut-off $L$ limiting the angular momentum $j$ .We then have the kernel

\begin{equation}
\begin{aligned}
&\left|j\right|\sigma_{j}\left(t,\rho,\phi,\theta\right)=
\frac{1}{2\pi }e^{ijt} \int_{0}^{2\pi} d\tau K^{LLM}_{L}\left(\rho,
\phi,\theta|\tau \right) e^{ij\tau}, \,\, \left|j\right|\leq L\\
&0= \frac{1}{2\pi }e^{ijt} \int_{0}^{2\pi} d\tau
K^{LLM}_{L}\left(\rho, \phi,\theta|\tau \right) e^{ij\tau}, \,\,
\left|j\right|\leq L, \,\, \left|j\right|> L.
\end{aligned}
\end{equation}
The kernel with the cutoff is given by
\begin{equation}
\begin{aligned}
&K^{LLM}_{L}\left(\rho, \phi,\theta|\tau \right)= K^{LLM}\left(\rho, \phi,\theta|\tau \right)+\\
&\frac{-\cos\left[L\left(\tau - \phi \right) \right]+a
\cos\left[\left(L-1\right)\left(\tau - \phi \right)\right]
}{1+a^{2}- 2a \cos\left(\tau-\phi \right)}a^{L}+\\
& \frac{aL \cos\left[\left(L+2\right)\left(\tau - \phi
\right)\right]-\left[1+L+2La^{2} \right]
\cos\left[\left(L+1\right)\left(\tau - \phi
\right)\right]}{\left[1+a^{2}- 2a \cos\left(\tau-\phi \right)
\right]^{2}}a^{L}+\\
& \frac{-a^{2}\left(L+1\right) \cos\left[\left(L-1\right)\left(\tau
- \phi \right)\right]+a\left[2+2L+La^{2} \right]
\cos\left[L\left(\tau - \phi \right)\right]}{\left[1+a^{2}- 2a
\cos\left(\tau-\phi \right) \right]^{2}}a^{L}
\end{aligned}
\end{equation}
We see that we have a strong convergence
\begin{equation}
\lim_{L\rightarrow\infty}K^{LLM}_{L}\left(\rho, \phi,\theta|\tau
\right)=K^{LLM}\left(\rho, \phi,\theta|\tau \right)
\end{equation}

\subsection{Correspondence with the 2d Black Hole}
To proceed with the construction of the kernel in our more general
two dimensional case it is also useful to take note of a correspondence
with an equivalent problem that was considered in the case of a 2d black hole.
We show in what follows that there is a simple connection
between  "off-shell" black hole wavefunctions and
on-shell AdS wavefunctions that we have identified.

The wavefunctions that we consider correspond to highest weight states on
$SO\left(4\right)$ but with a nontrivial dependence on the radial coordinate of AdS .

\begin{equation*}
f=\cos^{l}\left(\theta\right)e^{il\phi} \psi\left(t,\sigma \right)
\end{equation*}

We have the following eigenequatin for $\psi$ 

\begin{equation*}
\begin{aligned}
&-\cos^{2}\left(\sigma\right)\partial_{t}^{2}\psi
+\cos^{2}\left(\sigma\right)
\partial_{\sigma}^{2} +\cot\left(\sigma\right)\partial_{\sigma}
\psi= l\left(l-2\right)\psi \Rightarrow \\
&-\partial_{t}^{2}\psi+ \partial_{\sigma}^{2}\psi +
\frac{1}{\cos\left( \sigma\right) \sin\left(
\sigma\right)}\partial_{\sigma} \psi=
\frac{l\left(l-2\right)}{\cos^{2}\left(\sigma\right)} \psi
\end{aligned}
\end{equation*}

with the integration measure 
\begin{equation*}
dm=\sqrt{-g}g^{00}dt d\sigma=\tan\left(\sigma\right)d\sigma.
\end{equation*}
A change to a new function
\begin{equation*}
R= \frac{1}{\cos\left(\sigma\right)}\psi
\end{equation*}
with the new measure
\begin{equation*}
dm=\frac{1}{2} \sin\left(2\sigma \right) dt d\sigma
\end{equation*}
leads to the equation
\begin{equation*}
\begin{aligned}
&-\partial_{t}^{2}R+
\frac{1}{\cos\left(\sigma\right)}\left[\partial_{\sigma}^{2} +
\frac{1}{\cos\left( \sigma\right) \sin\left(
\sigma\right)}\partial_{\sigma}\right]\cos\left(\sigma\right)R=
\frac{l\left(l-2\right)}{\cos^{2}\left(\sigma\right)} R
\end{aligned}
\end{equation*}

or

\begin{equation*}
\begin{aligned}
& \partial_{\sigma}^{2}R+
2\cot\left(2\sigma\right)\partial_{\sigma}R=
\frac{l\left(l-2\right)+1}{\cos^{2}\left(\sigma\right)} R
-\omega^{2}R+R
\end{aligned}
\end{equation*}
This can be compared with the 2d black hole equation\cite{JY} defined as a coset
$\widetilde{SL}\left(2,R\right)/U\left(1\right)$. For the case of
the Lorentzian black hole they are specified by the eigenvalue
equation
\begin{equation*}
\begin{aligned}
\Delta_{0}T_{\nu}^{\lambda}&=\left(-\frac{1}{4}-\lambda^{2}\right)T_{\nu}^{\lambda} \Rightarrow\\
-\frac{1}{4\sinh
\left(\frac{r}{2}\right)}\partial_{\tau}^{2}T_{\nu}^{\lambda}
+\partial_{r}^{2}T_{\nu}^{\lambda}+
\coth\left(r\right)\partial_{r}T_{\nu}^{\lambda}&= \left(-\frac{1}{4}-\lambda^{2}\right)T_{\nu}^{\lambda} \Rightarrow\\
\frac{1}{\sinh \left(\frac{r}{2}\right)}\nu^{2}T_{\nu}^{\lambda}
+\partial_{r}^{2}T_{\nu}^{\lambda} +
\coth\left(r\right)\partial_{r}T_{\nu}^{\lambda}&=\left(-\frac{1}{4}-\lambda^{2}\right)T_{\nu}^{\lambda}
\end{aligned}
\end{equation*}

%\end{equation}
and the inner product is defined through the  integration
measure
\begin{equation*}
\langle
T_{\nu}^{\lambda}|T_{\nu^{\prime}}^{\lambda^{\prime}}\rangle=
\delta\left(\nu-\nu^{\prime} \right)\int_{0}^{\infty}dr
\sinh\left(r\right)
\left(T_{\nu}^{\lambda}\left(r\right)\right)^{*}
T_{\nu^{\prime}}^{\lambda^{\prime}}\left(r\right).
\end{equation*}

We see that the two problems are related,through the following transformation
transformations
\begin{equation*}
\begin{aligned}
& l\rightarrow 1-2i\nu\\
& \omega\rightarrow i2\lambda\\
& \sigma\rightarrow \frac{i}{2}\left(r+\pi\right)
\end{aligned}
\end{equation*}

In ref. \cite{JY} a transformation was constructed relating the wavefunctions in
the black hole case to those of a c=1 matrix model. The transformation reads
\
\begin{equation*}
T^{\lambda}_{\nu}= \int_{-\infty}^{+\infty} dt_{0}\int_{0}^{+\infty}
d\sigma
\delta\left[\sinh\left(\frac{r}{2}\right)\sinh\left(\frac{2t_{0}}{3}-\tau
\right)- \cosh\left(2\sigma\right) \right]
e^{-4i\frac{t_{0}}{3}}\cos\left(4\lambda\sigma \right).
\end{equation*}
and it involves a nontrivial kernel which specifies a canonical transformation
from one problem to another.
In the present case we will follow the construction of \cite{JY} and construct an analogous
kernel which will relate $AdS$ (and $S$) wavefunctions to those of the matrix eigenvalue problem.

\subsection{The AdS Kernel}

We first give the main formulas defining the AdS kernel. The wavefunctions obey the equations
\begin{equation}
\begin{aligned}
&\nabla^{2}_{S^{3}}\sigma_{j,n}\left(t,\rho,\phi,\theta\right)=
-\left|j\right|\left(\left|j\right|+2 \right)
\sigma_{j,n}\left(t,\rho,\phi,\theta\right)\\
&\nabla^{2}_{AdS^{3}}\sigma_{j,n}\left(t,\rho,\phi,\theta\right)=
\left|j\right|\left(\left|j\right|-2 \right)
\sigma_{j,n}\left(t,\rho,\phi,\theta\right)\\
&-i\frac{\partial}{\partial\phi}\sigma_{j,n}\left(t,\rho,\phi,\theta\right)=j
\sigma_{j,n}\left(t,\rho,\phi,\theta\right)
\end{aligned}
\end{equation}
and have an explicit solution in terms of hypergeometric functions
\begin{equation}
\begin{aligned}
&\sigma_{j,n}\left(t,\rho,\phi,\theta\right)=
e^{\frac{j}{\left|j\right|} i\omega_{j,n}
t}\cos^{\left|j\right|}\theta e^{\frac{j}{\left|j\right|}i j \phi}
\cosh^{-\left(\left|j\right|+2n \right)}\rho F\left(1-j-n, -n;1;
-\sinh^{2}\rho \right)\\
&\omega_{j,n}=\left|j\right|+2n
\end{aligned}
\end{equation}
We now use the integral representation
\begin{equation}
\begin{aligned}
&\sigma_{j,n}\left(t,\rho,\phi,\theta\right)=e^{\frac{j}{\left|j\right|} i\omega_{j,n} t}\\
& \oint_{C}dz\frac{1}{i2\pi z}
\left[\cosh\rho
e^{-\frac{j}{\left|j\right|}i\phi} \left( \cosh\rho + z\sinh\rho
\right)\right]^{-\left|j\right|-2n} \times \\
&\left[\left(\frac{\cosh\rho}{z}
+ \sinh\rho\right)\left(\cosh\rho +
z\sinh\rho\right)e^{-2\frac{j}{\left|j\right|}i\phi} \right]^{n}=\\
&e^{\frac{j}{\left|j\right|} i\omega_{j,n} t}
\oint_{C}dz\frac{1}{i2\pi
z}\left[[e^{\frac{j}{\left|j\right|}i\phi}w\right]^{\left|j\right|+2n}
\left[e^{-2\frac{j}{\left|j\right|}i\phi}v\right]^{n}
\end{aligned}
\end{equation}
where
\begin{equation}
\begin{aligned}
&w=\frac{1}{\cosh\rho \left(
\cosh\rho + z\sinh\rho \right)}\\
&v=\left(\frac{\cosh\rho}{z} + \sinh\rho\right)\left(\cosh\rho +
z\sinh\rho\right)
\end{aligned}
\end{equation}

\begin{equation}
\left|j\right|+2n\leq L
\end{equation}
to derive the kernel defined through
\begin{equation}
\begin{aligned}
&\left|j\right|\sigma_{j,n}=e^{\frac{j}{\left|j
\right|}i\omega_{j,n} t} \frac{1}{4\pi^{2}} \int_{0}^{2\pi}d\sigma
\int_{0}^{2\pi}d\tau K_{L}\left(\rho, \phi,\theta| \sigma,
\tau\right)e^{i\frac{j}{\left|j\right|}\left[\left(\left|j\right|+
2n \right)\tau+ n\sigma \right]},\\
&K_{L}\left(\rho, \phi,\theta| \sigma, \tau\right)=
\oint_{C}\frac{dz}{i2\pi z} \left(F_{L}\left(w|\tau\right)
G_{L}\left(v|\sigma\right)-2 \tilde{F}_{L}\left(w|\tau\right)
\tilde{G}_{L}\left(v|\sigma\right)\right).
\end{aligned}
\end{equation}
Explicitely, the functions involved in the definition of the kernel 
can be worked out after an introduction of a cut off L for convergence.
They take the slightly long forms:
\begin{equation}
\begin{aligned}
&F_{L}\left(w|\tau\right)= \frac{1-4w^{2}
-w^{4}+4w^{3}\cos\left(\tau-\phi \right)}{\left[1+w^{2}- 2w
\cos\left(\tau-\phi \right) \right]^{2}}+\\
&\frac{-\cos\left[L\left(\tau - \phi \right) \right]+w
\cos\left[\left(L-1\right)\left(\tau - \phi \right)\right]
}{1+w^{2}- 2w \cos\left(\tau-\phi \right)}w^{L}+\\
& \frac{wL \cos\left[\left(L+2\right)\left(\tau - \phi
\right)\right]-\left[1+L+2Lw^{2} \right]
\cos\left[\left(L+1\right)\left(\tau - \phi
\right)\right]}{\left[1+w^{2}- 2w \cos\left(\tau-\phi \right)
\right]^{2}}w^{L}+\\
& \frac{-w^{2}\left(L+1\right) \cos\left[\left(L-1\right)\left(\tau
- \phi \right)\right]+w\left[2+2L+Lw^{2} \right]
\cos\left[L\left(\tau - \phi \right)\right]}{\left[1+w^{2}- 2w
\cos\left(\tau-\phi \right) \right]^{2}}w^{L}
\end{aligned}
\end{equation}

\begin{equation}
\begin{aligned}
&G_{L}\left(v|\sigma\right)= \frac{1-v^{2} }{1+v^{2}- 2v
\cos\left(\sigma-2\phi \right)}+\\
&\frac{-\cos\left[L\left(\sigma - 2\phi \right) \right]+v
\cos\left[\left(L-1\right)\left(\sigma - 2\phi \right)\right]
}{1+v^{2}- 2v \cos\left(\sigma-2\phi \right)}v^{L}
\end{aligned}
\end{equation}

\begin{equation}
\begin{aligned}
&\tilde{F}_{L}\left(w|\tau\right)= \frac{1-w^{2} }{1+w^{2}- 2w
\cos\left(\tau-\phi \right)}+\\
&\frac{-\cos\left[L\left(\tau - \phi \right) \right]+w
\cos\left[\left(L-1\right)\left(\tau - \phi \right)\right]
}{1+w^{2}- 2w \cos\left(\tau-\phi \right)}w^{L}
\end{aligned}
\end{equation}

\begin{equation}
\begin{aligned}
&\tilde{G}_{L}\left(v|\sigma\right)= \frac{v\left( v^{2}+1
\right)\cos\left(\tau-2\phi \right)-2v^{2}}{\left[1+v^{2}- 2v
\cos\left(\tau-2\phi \right) \right]^{2}}+\\
& \frac{vL \cos\left[\left(L+2\right)\left(\tau - 2\phi
\right)\right]-\left[1+L+2Lv^{2} \right]
\cos\left[\left(L+1\right)\left(\tau - 2\phi
\right)\right]}{\left[1+v^{2}- 2v \cos\left(\tau-2\phi \right)
\right]^{2}}v^{L}+\\
& \frac{-v^{2}\left(L+1\right) \cos\left[\left(L-1\right)\left(\tau
- 2\phi \right)\right]+v\left[2+2L+Lv^{2} \right]
\cos\left[L\left(\tau - 2\phi \right)\right]}{\left[1+v^{2}- 2v
\cos\left(\tau-2\phi \right) \right]^{2}}v^{L}
\end{aligned}
\end{equation}

\subsection{The Sphere Kernel}

We now consider the second sequence of wavefunctions,which are characterized by a nontrivial dependence on the radial coordinate of the sphere. The wave equations read

\begin{equation}
\begin{aligned}
&\nabla^{2}_{S^{3}}\sigma_{j,n}\left(t,\rho,\phi,\theta\right)=
-\left(\left|j\right|+2n \right)\left(\left|j\right|+2n+2 \right)
\sigma_{j,n}\left(t,\rho,\phi,\theta\right)\\
&\nabla^{2}_{AdS^{3}}\sigma_{j,n}\left(t,\rho,\phi,\theta\right)=
\left(\left|j\right|+2n \right)\left(\left|j\right|+2n-2 \right)
\sigma_{j,n}\left(t,\rho,\phi,\theta\right)\\
&-i\frac{\partial}{\partial\phi}\sigma_{j,n}\left(t,\rho,\phi,\theta\right)=j
\sigma_{j,n}\left(t,\rho,\phi,\theta\right)
\end{aligned}
\end{equation}
In the coordinate system where the metric is

\begin{equation*}
ds^{2}=-\cosh^{2}\rho dt^{2} + d\rho^{2}+ \sinh^{2}\rho d\psi^{2}+
d\theta^{2} + \cos^{2}\theta d\phi^{2}+ \sin^{2}\theta
d\tilde{\psi}^{2},
\end{equation*}
the  normalizable solutions are given by
\begin{equation}
\begin{aligned}
&\sigma_{j,n}= e^{\frac{j}{\left|j \right|}i\omega_{j,n} t}
e^{ij\phi} \cosh^{-\left|j\right|-2n}\rho
\cos^{\left|j\right|}\theta F\left(1+\left|j\right|+n, -n;1;
\sin^{2}\theta \right)\\
&\omega_{j,n}=\left|j\right|+2n
\end{aligned}
\end{equation}
We use the integral form of the
wavefunctions
\begin{equation}
\begin{aligned}
\sigma_{j,n}&=e^{\frac{j}{\left|j \right|}i\omega_{j,n} t}
e^{ij\phi} \oint_{C}\frac{dz}{i2\pi z} \left(\frac{ \cos\theta -
z\sin\theta} {\cosh\rho} \right)^{\left|j\right|+2n}
\left(\frac{\cos\theta +
\frac{\sin\theta}{z}} {\cos\theta - z\sin\theta} \right)^{n}\\
&= e^{\frac{j}{\left|j \right|}i\omega_{j,n} t}
\oint_{C}dz\frac{dz}{i2\pi z} \left(\frac{ \cos\theta - z\sin\theta}
{\cosh\rho}
e^{i\frac{j}{\left|j\right|}\phi}\right)^{\left|j\right|+2n}
\left(\frac{\cos\theta + \frac{\sin\theta}{z}} {\cos\theta -
z\sin\theta}e^{-i2\frac{j}{\left|j\right|}\phi} \right)^{n}\\
&= e^{\frac{j}{\left|j \right|}i\omega_{j,n} t}
\oint_{C}dz\frac{dz}{i2\pi z}
\left[[e^{\frac{j}{\left|j\right|}i\phi}w\right]^{\left|j\right|+2n}
\left[e^{-2\frac{j}{\left|j\right|}i\phi}v\right]^{n}
\end{aligned}
\end{equation}
where $C$ is the unit circle on the complex plane and we defined

\begin{equation}
\begin{aligned}
w= \frac{ \cos\theta - z\sin\theta} {\cosh\rho}
\\
v= \frac{\cos\theta + \frac{\sin\theta}{z}} {\cos\theta -
z\sin\theta}
\end{aligned}
\end{equation}

Introducing a cut off on the angular momentum as we had
for the LLM case
\begin{equation}
\left|j\right|+2n\leq L
\end{equation}
we rewrite the wavefunction as
\begin{equation}
\begin{aligned}
&\left(\left|j\right|+2n\right)\sigma_{j,n}=e^{\frac{j}{\left|j
\right|}i\omega_{j,n} t} \frac{1}{4\pi^{2}} \int_{0}^{2\pi}d\sigma
\int_{0}^{2\pi}d\tau K_{L}\left(\rho, \phi,\theta| \sigma,
\tau\right)e^{i\frac{j}{\left|j\right|}\left[\left(\left|j\right|+
2n \right)\tau+ n\sigma \right]},\\
&K_{L}\left(\rho, \phi,\theta| \sigma, \tau\right)=
\oint_{C}\frac{dz}{i2\pi z} F_{L}\left(w|\tau\right)
G_{L}\left(v|\sigma\right).
\end{aligned}
\end{equation}
The functions $F_{L}\left(w|\tau\right)$ and
$G_{L}\left(v|\sigma\right)$ specifying the kernel in this case are found
to be given by
\begin{equation}
\begin{aligned}
&F_{L}\left(w|\tau\right)= \frac{1-4w^{2}
-w^{4}+4w^{3}\cos\left(\tau-\phi \right)}{\left[1+w^{2}- 2w
\cos\left(\tau-\phi \right) \right]^{2}}+\\
&\frac{-\cos\left[L\left(\tau - \phi \right) \right]+w
\cos\left[\left(L-1\right)\left(\tau - \phi \right)\right]
}{1+w^{2}- 2w \cos\left(\tau-\phi \right)}w^{L}+\\
& \frac{wL \cos\left[\left(L+2\right)\left(\tau - \phi
\right)\right]-\left[1+L+2Lw^{2} \right]
\cos\left[\left(L+1\right)\left(\tau - \phi
\right)\right]}{\left[1+w^{2}- 2w \cos\left(\tau-\phi \right)
\right]^{2}}w^{L}+\\
& \frac{-w^{2}\left(L+1\right) \cos\left[\left(L-1\right)\left(\tau
- \phi \right)\right]+w\left[2+2L+Lw^{2} \right]
\cos\left[L\left(\tau - \phi \right)\right]}{\left[1+w^{2}- 2w
\cos\left(\tau-\phi \right) \right]^{2}}w^{L}
\end{aligned}
\end{equation}
and
\begin{equation}
\begin{aligned}
&G_{L}\left(v|\sigma\right)= \frac{1-v^{2} }{1+v^{2}- 2v
\cos\left(\sigma-2\phi \right)}+\\
&\frac{-\cos\left[L\left(\sigma - 2\phi \right) \right]+v
\cos\left[\left(L-1\right)\left(\sigma - 2\phi \right)\right]
}{1+v^{2}- 2v \cos\left(\sigma-2\phi \right)}v^{L}
\end{aligned}
\end{equation}
Let us now make the following comment regarding the cutoff that we have used. Since it 
imposes an upper limit on angular momenta it clearly plays a role of the 'exclusion principle'.
Its removal seems to lead to singularities both in the sphere and the AdS case. One should
remember then that this analysis is done at the linearized level, 
so there is no essential difference between the two cases.
We can also  show that if we restrict our attention to $1/2$ BPS wavefunctions (which
would correspond to $n=0$), the above
kernel reduces to the kernel that we have found from the LLM construction. We notice that the function
$F_{L}\left(w|\tau\right)$ is analytic in the unit circle $C$ of the
$z-$plane for every value of the remaining variables.
\begin{equation}
\left|j\right|\sigma_{j,0}=e^{ij t} \frac{1}{4\pi^{2}}
\int_{0}^{2\pi}d\tau \int_{0}^{2\pi}d\sigma K_{L}\left(\rho,
\phi,\theta| \sigma, \tau\right)e^{ij\tau}
\end{equation}
Performing the integral over $\sigma$ gives
\begin{equation}
\int_{0}^{2\pi}d\sigma G_{L}\left(v|\sigma\right)=1
\end{equation}
which establishes the result
\begin{equation}
\int_{0}^{2\pi}d\sigma K_{L}\left(\rho, \phi,\theta| \sigma,
\tau\right)=\oint_{C}\frac{dz}{i2\pi z} F_{L}\left(w|\tau\right)=
\left.F_{L}\left(w|\tau\right)\right|_{z=0}=K^{LLM}_{L}\left(\rho,
\phi,\theta|\tau \right)
\end{equation}

\section{Conclusion}

We have in the present work considered the simple a complex two
matrix model with a purpose of developing further its
correspondence with AdS eigenstates. We develop a (hybrid) formalism to 
construct a two dimensional sequence of invariant matrix model
eigenstates. Here one of the (matrix) degrees of freedom is treated in
a density representation (in a manner analogous to the one matrix collective 
field theory), while the other is represented in the coherent state picture.
This leads to a sequence of (integral) equations which we
then solve for the case of the oscillator potential. The two dimensional set of eigenstates
extends the one dimensional space representing the eigenstates of free fermions.
As such this extension allows a nontrivial probe of one further extra dimension .
This as we argue can be mapped into either the radial coordinate of AdS or the 
radial coordinate of the sphere.

The mapping between states of the matrix model and the wavefunctions 
of SUGRA is one to one. As such it differs from the holographic map where 
one of the dimensions
is projected out. In the present case the map can be described by a (two dimensional) 
kernel in paralel
with similar maps in the case of 2d noncritical string theory. We also note that
leg factors of this kind were found in the pp-wave map of \cite{Dobashi:2004nm}.

When applied to a one
dimensional subspace of $1/2$ BPS wavefunctions our kernel reduces to the (linearized)
map of LLM. In the construction of the extended map one seemingly requires a 
cutoff providing an interesting implementation of the 'exclusion principle'. 
The understanding of this cutoff is clearly of further interest.

It should be commented that much like in the $1/2$ BPS case of  free fermions 
the model considered is that of simple decoupled harmonic oscillators. 
Yang-Mills type interactions present in the full theory might be of  relevance 
but are not included in our study. For the case of
$1/2$ BPS correlators there are theorems regarding the absence
of coupling constant corrections. It can be hoped that this will persist for the 
present set of states. Certainly, the effect of coupling constant correction
deserves to be investigated (e.g, \cite{deMelloKochPV}). 
It is also of interest to extend the present map to 
a still larger set of eigenstates.

\section{Acknowledgements}

One of us (J.P.R.) would like to thank the High Energy Group and the Physics Department of
Brown University for making possible his stay at Brown during part of his sabbatical, and for
the hospitality extended to him during this visit.

\newpage
\section{Appendices}

\subsection{Appendix A: Expanding the LLM solution in fluctuations}
In this section we would like to expand the circular droplet
solution in "off-shell" fluctuations of the matrix model and see the
equations of motion these fluctuations satisfy from the bosonic
equations of motion of gravity.This analysis was also performed independently in a 
recent paper \cite{Grant:2005qc}.

The general $\frac{1}{2}$BPS LLM solution for the metric is
determined by the function
\begin{eqnarray*}
u\left(x_{1},x_{2},y \right)= \frac{y^{2}}{\pi}\int d\tilde{x}^{2}
u\left(\tilde{x}_{1},\tilde{x}_{2},0 \right)
\frac{1}{\left[\left(\vec{x}-\tilde{\vec{x}}
\right)^{2}+y^{2}\right]^{2}}
\end{eqnarray*}
with $u\left(\tilde{x}_{1},\tilde{x}_{2},0 \right)$ being the phase
space distribution of the fermions in the matrix model picture.
Parametrizing the boundary of the fermi surface using the polar
coordinates representation
\begin{eqnarray*}
\tilde{x}^{2}_{1}\left(\phi,t\right)+\tilde{x}_{2}^{2}\left(
\phi,t\right)=
R_{AdS}^{2}+\sum_{n>0}p_{n}\left(t\right)\sin\left(n\phi\right) +
nq_{n}\left(t\right)\cos\left(n\phi\right)
\end{eqnarray*}
the phase space density becomes
\begin{eqnarray*}
u\left(r,\phi,0,t \right)=
-\theta\left(\sqrt{R_{AdS}^{4}+\sum_{n>0}p_{n}\left(t\right)\sin\left(n\phi\right)
+ nq_{n}\left(t\right)\cos\left(n\phi\right)}-r\right)+\frac{1}{2}
\end{eqnarray*}
and after approximating at first order in pertubations $p_{n},q_{n}$
the distribution becomes.
\begin{eqnarray*}
u\left(r,\phi,0,t \right) &\approx & \frac{1}{2}-\theta\left(R^{2}_{AdS}-
r \right) \\
&-& \delta\left(R^{2}_{AdS}-r
\right)\left[\sum_{n>0}\frac{p_{n}\left(t\right)}{2R^{2}_{AdS}}
\sin\left(n\phi\right) +
n\frac{q_{n}\left(t\right)}{2R^{2}_{AdS}}\cos\left(n\phi\right)
\right]
\end{eqnarray*}
The field that is produced is given then by
\begin{eqnarray*}
u\left(r,\phi,y,t \right)&=& u_{AdS}\left(r,\phi,y,t \right)+
\tilde{u}\left(r,\phi,y,t \right)\\
u\left(r,\phi,y,t \right)&=& u_{AdS}\left(r,\phi,y,t \right)
-\frac{y^{2}}{2\pi }\int_{0}^{2\pi}d\tilde{\phi}
\frac{\sum_{n>0}p_{n}\left(t\right) \sin\left(n\tilde{\phi}\right) +
nq_{n}\left(t\right)\cos\left(n\tilde{\phi}\right)}{\left[R_{AdS}^{4}+r^{2}+y^{2}-2r
R_{AdS}^{2} \cos\left(\tilde{\phi}-\phi\right) \right]^{2}}.
\end{eqnarray*}
The above integral can be computed from the more general
\begin{equation*}
\begin{aligned}
&\int_{0}^{2\pi}d\phi\frac{e^{im\phi}}{\left( a-2\cos\left(\phi
\right) \right)^{2}}=\frac{1}{i} \oint_{C}dz
\frac{z^{n+1}}{\left(z^{2}-az+1 \right)^{2}}= \frac{1}{i}
\oint_{C}dz \frac{z^{n+1}}{\left(z-z_{+} \right)^{2}\left(z-z_{-}
\right)^{2}}=\\
&2\pi \left. \frac{d}{dz} \frac{z^{n+1}}{\left(z-z_{+} \right)^{2}}
\right|_{z=z_{-}}= 2\pi
\frac{z_{-}^{n}}{\left(z_{-}-z_{+}\right)^{2}}\left[n+
\frac{z_{+}+z_{-}}{z_{+}-z_{-}} \right],\\
&z_{\pm}=\frac{a\pm\sqrt{a^{2}-4}}{2}.
\end{aligned}
\end{equation*}
Where the contour $C$ is the unit circle on the complex plane of
integration and we only picked the contribution from $z_{-}$ which
is the root that is inside the circle for $a>1$.

After setting
\begin{eqnarray*}
y&=&R^{2}_{AdS} \sinh\rho \sin\theta\\
r&=&R^{2}_{AdS} \cosh\rho \cos\theta
\end{eqnarray*}
the result is given by
\begin{equation*}
\begin{aligned}
&u\left(\rho,\phi,\theta,t \right)=u_{AdS}\left(\rho,\phi,\theta,t
\right)-
\frac{1}{R_{AdS}^{4}}\frac{\sinh^{2}\rho
\sin^{2}\theta}{\left(\cosh^{2}\rho -\cos^{2}\theta\right)^{2} } \times \\
&\sum_{n>0}\left(\frac{\cos\theta}{\cosh\rho}
\right)^{n}\left[n+\frac{\cosh^{2}\rho
+\cos^{2}\theta}{\cosh^{2}\rho -\cos^{2}\theta} \right]\left[
p_{n}\left(t\right)\sin\left(n\phi\right) +
nq_{n}\left(t\right)\cos\left(n\phi\right)\right]\\
&\text{where}\\
& u_{AdS}\left(\rho,\phi,\theta,t \right)= \frac{1}{2}
\frac{\sinh^{2}\rho-\sin^{2}\theta}{\sinh^{2}\rho+ \sin^{2}\theta }.
\end{aligned}
\end{equation*}
%Repeating the same steps for the fluctuations in the vector field we
%obtain
%\begin{equation*}
%\begin{aligned}
%&V_{\phi}\left(\rho,\phi,\theta,t \right)=\\
%&V_{\phi}^{AdS}\left(\rho,\phi,\theta,t \right)+
%\tilde{V}_{\phi}\left(\rho,\phi,\theta,t \right)=\\
%&V_{\phi}^{AdS}\left(\rho,\phi,\theta,t \right)+\\
%&\frac{1}{4}\frac{1}{\left(\cosh^{2}\rho - \cos^{2}\theta
%\right)\cosh\rho \cos\theta}
%\sum_{n>0}np_{n}\left(\frac{\cos\theta}{\cosh\rho}\right)^{n}\sin\left(n\phi\right)
%\left[\left(m+1\right)\cos^{2}\theta+\left(m-1\right)\cosh^{2}\rho+
%\left(\frac{\left(\cosh^{2}\rho +\cos^{2}\theta
%\right)^{2}}{\cosh^{2}\rho
%-\cos^{2}\theta} \right)^{2}\right]+\\
%&+\frac{1}{4}\frac{1}{\left(\cosh^{2}\rho - \cos^{2}\theta
%\right)\cosh\rho \cos\theta} \sum_{n>0}n
%^{2}q_{n}\left(\frac{\cos\theta}{\cosh\rho}\right)^{n}\cos\left(n\phi\right)
%\left[\left(m+1\right)\cos^{2}\theta-\left(m-1\right)\cosh^{2}\rho+
%\left(\cosh^{2}\rho +\cos^{2}\theta \right)^{2}\right]
%\end{aligned}
%\end{equation*}
%\begin{eqnarray*}
%V_{r}\left(\rho,\phi,\theta,t \right)=
%\tilde{V}_{r}\left(\rho,\phi,\theta,t \right)=
%\end{eqnarray*}

The perturbation of the metric on $S^{5}$ is given by
\begin{equation*}
\begin{aligned}
&\frac{1}{R^{2}_{AdS}}d\tilde{s}_{S^{5}}^{2}=\\
&-\frac{2}{\sinh\rho \sin\theta} \left(\cosh^{2}\rho \sin^{2}\theta
+ \sinh^{2}\rho \cos^{2}\theta
\right)\frac{u_{AdS}}{\sqrt{1-4u_{AdS}^{2}}}\tilde{u} d\theta^{2}\\
& \frac{4 \cosh\rho \sinh\rho \sin^{2}\theta V_{\phi}^{AdS
}}{\sqrt{1-4u_{AdS}^{2}}}\tilde{V}_{r} d\theta d\phi\\
&- \left[\frac{4 \sinh\rho \sin\theta
V_{\phi}^{AdS}}{\sqrt{1-4u_{AdS}^{2}}} \tilde{V}_{\phi} +
\frac{8\sinh\rho \sin\theta
u_{AdS}{V_{\phi}^{AdS}}^{2}}{\sqrt{1-4u_{AdS}^{2}}^{3}} \tilde{u} +
\frac{2\cosh^{2}\rho \cos^{2}\theta u_{AdS}}{\sinh\rho \sin\theta
\sqrt{1-4u_{AdS}^{2}}} \tilde{u}\right] d\phi^{2}\\
&-2 \sinh\rho \sin\theta \sqrt{\frac{1+2u_{AdS}}{1-2u_{AdS}}}
\frac{1}{\left(1+2u_{AdS}\right)^{2}}\tilde{u}
d\tilde{\Omega}_{3}^{2}.
\end{aligned}
\end{equation*}

At this point we would like to show that the degrees of freedom
$q_{n},p_{n}$ turn on the chiral primary fields $\sigma^{I}$ of IIB
SUGRA on $AdS_{5}\times S^{5}$. After performing the field dependent
coordinate transformation
\begin{equation*}
\begin{aligned}
&\theta\rightarrow \theta - \frac{1}{2R^{4}_{AdS}}\frac{\sin\theta
\cos\theta}{\cosh^{2}\rho-\cos^{2}\theta}
\sum_{n>0}\left(\frac{\cos\theta}{\cosh\rho} \right)^{n}\left[
p_{n}\left(t\right)\sin\left(n\phi\right) +
nq_{n}\left(t\right)\cos\left(n\phi\right)\right]\\
&\rho\rightarrow \rho+\frac{1}{R^{4}_{AdS}}\frac{\cos^{2}\theta \tanh\rho}{\cosh^{2}\rho-\cos^{2}\theta}
\sum_{n>0}\left(\frac{\cos\theta}{\cosh\rho} \right)^{n}\left[
p_{n}\left(t\right)\sin\left(n\phi\right) +
nq_{n}\left(t\right)\cos\left(n\phi\right)\right]\\
&t\rightarrow t +\frac{1}{R^{4}_{AdS}}
\sum_{n>0}\left(\frac{\cos\theta}{\cosh\rho} \right)^{n}\left[
p_{n}\left(t\right)\cos\left(n\phi\right)
-q_{n}\left(t\right)\sin\left(n\phi\right)\right]
\end{aligned}
\end{equation*}
the $\theta\theta$ and $\tilde{S}_{3}$ components of the first order
perturbed metric are scaled by
\begin{equation*}
\frac{2}{R^{4}_{AdS}}\left(n+1\right)
\sum_{n>0}\left(\frac{\cos\theta}{\cosh\rho} \right)^{n}\left[
p_{n}\left(t\right)\sin\left(n\phi\right) +
nq_{n}\left(t\right)\cos\left(n\phi\right)\right].
\end{equation*}
After this observation we may identify the chiral primary fields
\begin{equation*}
\sigma^{\pm n}=\frac{1}{8R^{4}_{AdS}}\frac{n+1}{n} \left[nq_{n}\mp i
p_{n}\right] \left(\frac{1}{\cosh\rho} \right)^{n}.
\end{equation*}
The correctly normalized action for the chiral primaries as given by
Seiberg et al. reads
\begin{equation*}
S=\sum_{n}\frac{8R^{8}_{AdS} n \left(n-1
\right)}{\left(n+1\right)^{2}}
\int_{AdS^{5}}dx^{5}\sqrt{g_{AdS^{5}}}\left[\sigma^{-n}\Box
 \sigma^{+n} - n\left(n-4 \right)\sigma^{-n} \sigma^{+n} \right]
\end{equation*}
which after performing the spatial integral on $AdS_{5}$ gives
\begin{equation*}
S=\sum_{n>0}\frac{1}{2}\int dt \left[\frac{1}{n^2}\dot{p}_{n}^{2} +
\dot{q}^{2}_{n}-n^{2}q_{n}^{2}-p_{n}^{2} \right]
\end{equation*}
and for each $n$ we have a four dimensional phase space.
Supersymmetry requires that
$\left(\partial_{t}-\partial_{\phi}\right)\sigma=0$ which for our
$0+1$ dimensional variables means that the "angular momentum" is
equal to the energy. Choosing an opposite chirality for the fermions
we would have had the condition
$\left(\partial_{t}+\partial_{\phi}\right)\sigma=0$ which would flip
the sign in the relation between energy and "angular momentum".

\subsection{Appendix B: Hermiticity and the zero impurity sector }

\noindent
The zero impurity sector is the usual single matrix problem for $M_{ij}$. We are interested in fluctuations
about this single matrix background. As is now well known \cite{JevickiMB},\cite{DasKA},\cite{DemeterfiCW},
this background is only exhibited as
the stationary point of an explicitly hermitean effective potential. We recall the construction of this
effective hamiltonian\cite{JevickiMB}.

\noindent
In order to take into account the non trivial Jacobian $J$ involved in the change from the original
variables to loop variables, one needs to implement the similarity transformation ($i$ is a generic loop variable)

$$
       \partial_i \to J^{1\over 2} \partial_i J^{-{1\over 2}} = \partial_i - {1\over 2}\partial_i \ln J
$$

\noindent
The Jacobian satisfies \cite{JevickiMB}

$$
                   \Omega_{ij} \partial_j \ln J = \omega_i - \partial_j \Omega_{ji}
$$

\noindent
The terms of the kinetic energy operator that are sufficient to generate the background and fluctuations are then
\cite{AffleckRG},\cite{RodriguesFS},\cite{JevickiHB}

\be\label{Terms}
   {-{1 \over 2}}\partial_i  \Omega_{ij} \partial_j + {1\over 8} \omega_i  \Omega^{-1}_{ij}    \omega_j
\ee

\noindent
In the zero impurity sector,

\bea
\omega(k,0) &=& - k \int_0^{k} dk' \psi(k',0) \psi(k-k',0) \nonumber\\
\Omega (k,0;k',0) &= & - k k' \psi(k+k',0)
\eea

\noindent
The $x$ representation of $\psi(k,0)$ is the usual density of eigenvalues:

$$
\psi(x,0)= \Sigma_{i} \delta (x - \lambda_i) ,
$$

\noindent
and

\beas
\Omega(x,0;y,0) &= &\partial_x \partial_y (\psi (x,0) \delta (x-y)) \\
\omega(x,0) &=& - 2 \partial_x \Big( \psi(x,0)  \int dz {\psi(z,0) \over x-z} \Big)
\eeas

\noindent
From (\ref{Terms}) we then obtain the form of the effective hamiltonian which is sufficient for the discussion of background
generation and fluctuations:

\beas
H &= &\int dx \int dy \Big( {-{1 \over 2}}  {\partial \over \partial \psi(x,0)} \Omega(x,0;y,0) {\partial \over \partial \psi(y,0)}
+ {1\over 8} \omega (x,0) \Omega^{-1} (x,0;y,0) \omega (y,0) \Big) \\
&+& \int dx \psi(x,0)({x^2 \over 2}- \mu)
   \Big)
\eeas

\noindent
where the Lagrange multiplier $\mu$ enforces the contraint

\be \label{Const}
\int dx \psi(x,0) = N .
\ee

\noindent
Since

$$
\partial_x \partial_y \Omega^{-1}(x,0;y,0) = {\delta(x-y)\over \psi(x,0)}
$$

\noindent
and

\be
\int dx \psi(x,0) \big( \int dy {\psi(y,0) \over x-y} \big)^2 = {\pi^2 \over 3} \int dx \psi^3(x,0),
\ee

\noindent
the effective hamiltonian becomes:

\be
{-{1 \over 2}}
\int dx \partial_x {\partial \over \partial \psi(x,0)} \psi(x,0) \partial_x {\partial \over \partial \psi(x,0)}
+ \int dx \Big( {\pi^2 \over 6}\psi^3(x,0) +  \psi(x,0)({x^2 \over 2}- \mu) \Big)
\ee

\noindent
To exhibit explicitely the $N$ dependence, we rescale

\bea \label{Rescaling}
x & & \to \sqrt{N} x \nonumber\\
\psi(x,0) & & \to \sqrt{N} \psi(x,0) \nonumber\\
-i {\partial \over \partial \psi(x,0)}\equiv \Pi(x) & & \to {1 \over N} \Pi(x) \\
\mu & &  \to N \mu\ \nonumber
\eea

\noindent
and obtain

\be 
H_{eff}^{0}= {{1 \over 2N^2}}
\int dx \partial_x \Pi(x) \psi(x,0) \partial_x \Pi(x) +
N^2 \Big(  \int dx {\pi^2 \over 6}\psi^3(x,0) +  \psi(x,0)({x^2 \over 2}- \mu) \Big),
\ee

\noindent
which is equation (\ref{HEffZero}) in the main text. 

\subsection{Appendix C: Marchesini-Onofri Kernel}

\noindent
We consider the problem of finding the spectrum of the operator

\begin{eqnarray}\label{Starting}
\int_{-\sqrt{2}}^{\sqrt{2}} dy { \phi_0(y) \over (x-y)^2}
\Big( f(x)- f(y) \Big) &=  \Big( -{d \over dx} \int_{-\sqrt{2}}^{\sqrt{2}} dy { \phi_0(y) \over (x-y)} \Big) f(x) 
\nonumber \\ &+
{d \over dx} \int_{-\sqrt{2}}^{\sqrt{2}} dy { \phi_0(y) f(y) \over (x-y)}
\end{eqnarray}

\noindent
We start with the second term and consider the following integral, in "time of flight" coordinates:

$$
\int_{-\pi}^{\pi} {dq \over \pi} {\pi \phi_0(q)} {e^{inq}\over x(q_0)-x(q)}, \quad n > 0
$$

\noindent
Note that the range of the integral extends over a full period $2L=2\pi$ of the classical motion. Therefore, the
integral above can be calculated by the residue theorem, by choosing a vertical path from $-\pi + i \infty$ to 
$-\pi$, then along the real axis from $-\pi$ to $\pi$, and then along a vertical path from $\pi$ to $\pi + i \infty$,
"closing" at $+ i \infty$. The contribution fom the vertical paths cancel, due to the periodicity of the classical
motion. The origin of the "time of flight" can always be chosen so that the only poles on the real axis occur at
$q=q_0$ and $q=-q_0$, corresponding to an even (in $q$) "displacement" $x(q)$ and odd "velocity" $\pi \phi_0(q)$.
We alwyas choose a principal value prescription for poles on the real axis (half of the residue). If there are no
other poles, as is the case in general for stabilised potentials, we obtain the result:

$$
\int_{-\pi}^{\pi} {dq \over \pi} {\pi \phi_0(q)} {e^{inq}\over x(q_0)-x(q)} =
2 i \int_{0}^{\pi} {dq \over \pi} {\pi \phi_0(q)} {\sin(nq) \over x(q_0)-x(q)} = - 2 i \cos(nq_0)
$$    
  
\noindent
In other words

\be\label{Qspace}
\int_{0}^{\pi} {dq \over \pi} {\pi \phi_0(q)} {\sin(nq) \over x(q_0)-x(q)} = - \cos(nq_0)
\ee

\noindent
Therefore

$$
\partial_q \int {dq' \over \pi} {\sin(nq') \over x(q)-x(q')} 
\equiv - i |\partial_q| ( \sin(nq) ) = n ( \sin(nq) )
$$

\noindent
In $x$ space,

$$ 
 \int_{-\sqrt{2}}^{\sqrt{2}} dy { \sin(nq(y)) \over (x-y)} = - \cos(nq(x))
$$

\noindent
It follows that the eigenvalue equation 

$$
{d \over dx} \int_{-\sqrt{2}}^{\sqrt{2}} {dy\over \pi} { \pi\phi_0(y) f_n (y) \over (x-y)} = \epsilon_n f_n
$$
 
\noindent
has solutions

$$
f_n (x)= {\sin(nq(x)) \over \pi\phi_0} = {\sin(nq(x)) \over {\sqrt{2}} \sin(q(x))}, 
\quad x(q) = - {\sqrt{2}} \cos(q), \quad \epsilon_n = n
$$
  
\noindent
This follows from the observation that in terms of time of flight coordinates, 
the above spectrum equation takes the form

$$
\partial_q \int {dq' \over \pi} {\pi \phi_0 (q') f_n (q') \over x(q)-x(q')} 
\equiv - i |\partial_q| ( \pi \phi_0 (q) f_n (q) )
$$

\noindent
Concerning the first term in (\ref{Starting}), we have already seen for the main text that it can be
obtained straightforwardly from the result (eqn. (\ref{BIPZ}))   

\be 
\int dz {\phi_0(z) \over (x-z)}=x.
\ee

\noindent
This equation is solved by the well known methods of ref. \cite{BrezinSV}. We point out that first term 
of (\ref{Starting}) can also be obtained in general by considering the integral

$$
\int_{-\pi}^{\pi} {dq \over \pi} {(\pi \phi_0(q))^2 \over (x(q_0)-x(q))^2} 
$$

\noindent
along the contour described above. There is now a contribution from "infinity", and one obtains the result
that the above integral equals $-1$

\subsection{Appendix D: Three impurities}

For three impurities, we have

\bea
\Psi(k,3)&=& \int_0^k dk_2 \int_0^{k_2} dk_1 Tr(B e^{i k_1 M} B e^{i (k_2-k_1) M} B e^{i (k-k_2) M} )\\
 &= &-3 \Sigma_{i,j,k}(V^{+}B V)_{ij} (V^{+}B V)_{jk} (V^{+}B V)_{ki}
{e^{ik\lambda_j}\over (\lambda_j-\lambda_i)(\lambda_j-\lambda_k)} \nonumber
\eea

\noindent
and

\be
\psi(x,3) = - 3 \Sigma_{i,j,k}(V^{+}B V)_{ij} (V^{+}B V)_{jk} (V^{+}B V)_{ki}
{\delta(x-\lambda_j) \over (x-\lambda_i)(x-\lambda_k)}
\ee

\noindent
After some algebra, one obtains

\beas
\Omega(k_0,0:k,3)
&=& - k_0 \Sigma_{i,j,k}(V^{+}B V)_{ij} (V^{+}B V)_{jk} (V^{+}B V)_{ki} \Big[
{-3 k e^{i(k+k_0)\lambda_i} \over (\lambda_i-\lambda_j)(\lambda_i-\lambda_k)} \\
&-& 3 i  e^{i(k+k_0)\lambda_i}\Big(
{1 \over (\lambda_i-\lambda_k)^2 (\lambda_i-\lambda_j)} +
{1 \over (\lambda_i-\lambda_j)^2 (\lambda_i-\lambda_k)}\Big)\\
&+& {3 i  e^{i k_0 \lambda_j} e^{i k \lambda_i} \over (\lambda_j-\lambda_i)^2 (\lambda_i-\lambda_k)}
+ {3 i  e^{i k_0 \lambda_k} e^{i k \lambda_i} \over (\lambda_k-\lambda_i)^2 (\lambda_i-\lambda_j)}
\Big]
\eeas

\noindent
and

\beas
\Omega(x,0:y,3)&= &\Sigma_{i,j,k}(V^{+}B V)_{ij} (V^{+}B V)_{jk} (V^{+}B V)_{ki} \times \\
\Big[
&-&3 \partial_x \partial_y \Big( \delta(x-y) {\delta(y-\lambda_i) \over (y - \lambda_j)
(y - \lambda_k)} \Big) \\
&-&3 \partial_x \Big(
\delta(x-\lambda_i) \delta(y-\lambda_i) \big(
{1 \over (y - \lambda_k)^2} {1 \over (y - \lambda_j)}
{1 \over (y - \lambda_j)^2} {1 \over (y - \lambda_k)}
\big)\Big)\\
&+&3 \partial_x \big(
{\delta(x-\lambda_j) \delta(y-\lambda_i)\over (\lambda_j - y)^2 (y - \lambda_k) }
\big)
+3 \partial_x \big(
{\delta(x-\lambda_k) \delta(y-\lambda_i)\over (\lambda_k - y)^2 (y - \lambda_j) }
\big)
\Big]
\eeas

\noindent
The $\Omega(x,0:y,3)$ term in (\ref{HTwoOme}) takes the form

\beas
& & {1\over 2} \int dx \int dy \Omega(x,0:y,3)
{\partial \ln J \over \partial \psi(x,0)} {\partial \over \partial \psi(y,3)} \\
& =&- {3\over 2} \Sigma_{i,j,k}(V^{+}B V)_{ij} (V^{+}B V)_{jk} (V^{+}B V)_{ki} \int dx \int dy \times \\
& & \Big[
\delta(x-y) {\delta(y-\lambda_i) \over (y - \lambda_j)
(y - \lambda_k)} \partial_x {\partial \ln J \over \partial \psi(x,0)}
\partial_y {\partial \over \partial \psi(y,3)}\\
&-& \delta(x-\lambda_i) \delta(y-\lambda_i) \big(
{1 \over (y - \lambda_k)^2} {1 \over (y - \lambda_j)}
{1 \over (y - \lambda_j)^2} {1 \over (y - \lambda_k)}
\big)\partial_x {\partial \ln J \over \partial \psi(x,0)}
{\partial \over \partial \psi(y,3)}\\
& & \big(
{\delta(x-\lambda_j) \delta(y-\lambda_i)\over (\lambda_j - y)^2 (y - \lambda_k) } +
{\delta(x-\lambda_k) \delta(y-\lambda_i)\over (\lambda_k - y)^2 (y - \lambda_j) }
\big)
\partial_x {\partial \ln J \over \partial \psi(x,0)}
{\partial \over \partial \psi(y,3)}\Big]
\eeas

\noindent
For the harmonic oscillator potential, we use the results (\ref{Jac})  and  (\ref{BIPZ}) , so that we can write

$$
\partial_x {\partial \ln J \over \partial \psi(x,0)} =
2 \int dz {\phi_0 (z) \over (x-z)} = 2 z .
$$

\noindent
Then

\beas
& & {1\over 2}\int dx \int dy \Omega(x,0:y,3)
{\partial \ln J \over \partial \psi(x,0)} {\partial \over \partial \psi(y,3)} = \\
&-& {3} \int dx  \Sigma_{i,j,k}(V^{+}B V)_{ij} (V^{+}B V)_{jk} (V^{+}B V)_{ki}
{\delta(x-\lambda_i) \over (x - \lambda_j) (x - \lambda_k)}
\int dz {\phi_0 (z) \over (x-z)}
\partial_x {\partial \over \partial \psi(x,3)}\\
& + & 3 \int dy  \Sigma_{i,j,k}(V^{+}B V)_{ij} (V^{+}B V)_{jk} (V^{+}B V)_{ki} \delta(y-\lambda_i)
{\partial \over \partial \psi(y,3)}\\
& & \big(
{\lambda_i \over (y - \lambda_k)^2} {1 \over (y - \lambda_j)} +
{\lambda_i \over (y - \lambda_j)^2} {1 \over (y - \lambda_k)}
-{\lambda_j\over (\lambda_j - y)^2 (y - \lambda_k) }
-{\lambda_k\over (\lambda_k - y)^2 (y - \lambda_j) }
\big)\\
&=&\int dx \int dz  {\phi_0 (z) \over (x-z)} \psi(x,3)\partial_x {\partial \over \partial \psi(x,3)}
- 2 \int dx \psi(x,3)\partial_x {\partial \over \partial \psi(x,3)}
\eeas

\noindent
Again, the first term above cancels the similar term in (\ref{HTwoOme}) , and we obtain
for the quadratic hamiltonian in the $3$ impurity sector:

\be
H_2^{s=3} = \int dx \int dz {\phi_0(z)\psi(x,3)- \psi(z,3) \phi_0(x) \over (x-z)^2} {\partial \over \partial \psi(x,3)}
- 2 \int dx \psi(x,3) {\partial \over \partial \psi(x,3)}
\ee

\noindent
This is again a shifted Marchesini-Onofri operator. The spectrum and eigenfunctions of this operator are

$$
      w_n=n-3 \quad ; \qquad \phi_n^{s=3}= {\sin (nq) \over \sqrt{2} \sin(q)} \quad ; \quad n=1,2,...
$$

\noindent
Adding the contribution from the $Tr (B {\partial / \partial B}) $ term of the Hamiltonian we obtain

\be
w_n= n \quad ; \qquad \phi_n^{s=3}= {\sin (nq) \over \sqrt{2} \sin(q)} \quad ; \quad n=1,2,...
\ee


\begin{thebibliography}{99}

%\cite{McGreevy:2000cw}
\bibitem{McGreevy:2000cw}
  J.~McGreevy, L.~Susskind and N.~Toumbas,
   ``Invasion of the giant gravitons from anti-de Sitter space,''
  %
  JHEP {\bf 0006}, 008 (2000)
  [arXiv:hep-th/0003075].
  %%CITATION = HEP-TH 0003075;%%



%\cite{Grisaru:2000zn}
\bibitem{Grisaru:2000zn}
  M.~T.~Grisaru, R.~C.~Myers and O.~Tafjord,
   ``SUSY and Goliath,''
  %
  JHEP {\bf 0008}, 040 (2000)
  [arXiv:hep-th/0008015].
  %%CITATION = HEP-TH 0008015;%%

%\cite{Hashimoto:2000zp}
\bibitem{Hashimoto:2000zp}
  A.~Hashimoto, S.~Hirano and N.~Itzhaki,
   ``Large branes in AdS and their field theory dual,''
  %
  JHEP {\bf 0008}, 051 (2000)
  [arXiv:hep-th/0008016].
  %%CITATION = HEP-TH 0008016;%%
%\cite{Das:2000st}
\bibitem{Das:2000st}
  S.~R.~Das, A.~Jevicki and S.~D.~Mathur,
   ``Vibration modes of giant gravitons,''
  %
  Phys.\ Rev.\ D {\bf 63}, 024013 (2001)
  [arXiv:hep-th/0009019].
  %%CITATION = HEP-TH 0009019;%%
  S.~R.~Das, A.~Jevicki and S.~D.~Mathur,
   ``Giant gravitons, BPS bounds and noncommutativity,''
  %
  Phys.\ Rev.\ D {\bf 63}, 044001 (2001)
  [arXiv:hep-th/0008088].
  %%CITATION = HEP-TH 0008088;%%

%\cite{Myers:2003bw}
\bibitem{Myers:2003bw}
  R.~C.~Myers,
  ``Nonabelian phenomena on D-branes,''
  Class.\ Quant.\ Grav.\  {\bf 20}, S347 (2003)
  [arXiv:hep-th/0303072].
  %%CITATION = HEP-TH 0303072;%%



%\cite{Bena:2004qv}
\bibitem{Bena:2004qv}
  I.~Bena and D.~J.~Smith,
  ``Towards the solution to the giant graviton puzzle,''
  Phys.\ Rev.\ D {\bf 71}, 025005 (2005)
  [arXiv:hep-th/0401173].
  %%CITATION = HEP-TH 0401173;%%

%\cite{Balasubramanian:2001nh}
\bibitem{Balasubramanian:2001nh}
  V.~Balasubramanian, M.~Berkooz, A.~Naqvi and M.~J.~Strassler,
   ``Giant gravitons in conformal field theory,''
   JHEP {\bf 0204}, 034 (2002)
  [arXiv:hep-th/0107119].
  %%CITATION = HEP-TH 0107119;%%

%\cite{Corley:2001zk}
\bibitem{Corley:2001zk}
  S.~Corley, A.~Jevicki and S.~Ramgoolam,
  ``Exact correlators of giant gravitons from dual N = 4 SYM theory,''
  Adv.\ Theor.\ Math.\ Phys.\  {\bf 5}, 809 (2002)
  [arXiv:hep-th/0111222].
  %%CITATION = HEP-TH 0111222;%%




%\cite{Berenstein:2004kk}
\bibitem{Berenstein:2004kk}
  D.~Berenstein,
  ``A toy model for the AdS/CFT correspondence,''
  JHEP {\bf 0407}, 018 (2004)
  [arXiv:hep-th/0403110].
  %%CITATION = HEP-TH 0403110;%%



%\cite{Lin:2004nb}
\bibitem{Lin:2004nb}
  H.~Lin, O.~Lunin and J.~Maldacena,
  ``Bubbling AdS space and 1/2 BPS geometries,''
  JHEP {\bf 0410}, 025 (2004)
  [arXiv:hep-th/0409174].
  %%CITATION = HEP-TH 0409174;%%

%\cite{deMelloKoch:2004ws}
\bibitem{deMelloKoch:2004ws}
  R.~de Mello Koch and R.~Gwyn,
  ``Giant graviton correlators from dual SU(N) super Yang-Mills theory,''
  JHEP {\bf 0411}, 081 (2004)
  [arXiv:hep-th/0410236].
  %%CITATION = HEP-TH 0410236;%%

%\cite{Gubser:2004xx}
\bibitem{Gubser:2004xx}
  S.~S.~Gubser and J.~J.~Heckman,
  ``Thermodynamics of R-charged black holes in AdS(5) from effective
  strings,''
  JHEP {\bf 0411}, 052 (2004)
  [arXiv:hep-th/0411001].
  %%CITATION = HEP-TH 0411001;%%

%\cite{Suryanarayana:2004ig}
\bibitem{Suryanarayana:2004ig}
  N.~V.~Suryanarayana,
  ``Half-BPS giants, free fermions and microstates of superstars,''
  arXiv:hep-th/0411145.
  %%CITATION = HEP-TH 0411145;%%

%\cite{Balasubramanian:2004nb}
\bibitem{Balasubramanian:2004nb}
  V.~Balasubramanian, D.~Berenstein, B.~Feng and M.~x.~Huang,
  ``D-branes in Yang-Mills theory and emergent gauge symmetry,''
  JHEP {\bf 0503}, 006 (2005)
  [arXiv:hep-th/0411205].
  %%CITATION = HEP-TH 0411205;%%

%\cite{Zamaklar:2004ak}
\bibitem{Zamaklar:2004ak}
  M.~Zamaklar and K.~Peeters,
  ``Holographic dynamics of unstable branes in AdS,''
  Comptes Rendus Physique {\bf 5}, 1071 (2004)
  [arXiv:hep-th/0412089].
  %%CITATION = HEP-TH 0412089;%%
%cite{Vaman}
\bibitem{Vaman}
 J.T. Liu, D.Vaman, W.Y. Wen (Michigan U., MCTP),. MCTP-04-67, Dec 2004. 31pp.
[arXiv: hep-th/0412043].
%%CITATION = HEP-TH 0412043;%%

%\cite{Martelli:2004xq}
\bibitem{Martelli:2004xq}
  D.~Martelli and J.~F.~Morales,
  ``Bubbling AdS(3),''
  JHEP {\bf 0502}, 048 (2005)
  [arXiv:hep-th/0412136].
  %%CITATION = HEP-TH 0412136;%%

%\cite{Sheikh-Jabbari:2005mf}
\bibitem{Sheikh-Jabbari:2005mf}
  M.~M.~Sheikh-Jabbari and M.~Torabian,
  ``Classification of all 1/2 BPS solutions of the tiny graviton matrix
  theory,''
  JHEP {\bf 0504}, 001 (2005)
  [arXiv:hep-th/0501001].
  %%CITATION = HEP-TH 0501001;%%

%\cite{Berenstein:2005vf}
\bibitem{Berenstein:2005vf}
  D.~Berenstein and S.~E.~Vazquez,
  ``Integrable open spin chains from giant gravitons,''
  arXiv:hep-th/0501078.
  %%CITATION = HEP-TH 0501078;%%

%\cite{Mandal:2005wv}
\bibitem{Mandal:2005wv}
  G.~Mandal,
  ``Fermions from half-BPS supergravity,''
  arXiv:hep-th/0502104.
  %%CITATION = HEP-TH 0502104;%%

%\cite{Horava:2005pv}
\bibitem{Horava:2005pv}
  P.~Horava and P.~G.~Shepard,
  ``Topology changing transitions in bubbling geometries,''
  JHEP {\bf 0502}, 063 (2005)
  [arXiv:hep-th/0502127].
  %%CITATION = HEP-TH 0502127;%%

%\cite{Berenstein:2005fa}
\bibitem{Berenstein:2005fa}
  D.~Berenstein, D.~H.~Correa and S.~E.~Vazquez,
  ``Quantizing open spin chains with variable length: An example from giant
  gravitons,''
  arXiv:hep-th/0502172.
  %%CITATION = HEP-TH 0502172;%%

%\cite{Takayanagi:2005tq}
\bibitem{Takayanagi:2005tq}
  T.~Takayanagi and S.~Terashima,
  ``c = 1 matrix model from string field theory,''
  arXiv:hep-th/0503184.
  %%CITATION = HEP-TH 0503184;%%

%\cite{Bak:2005ef}
\bibitem{Bak:2005ef}
  D.~Bak, S.~Siwach and H.~U.~Yee,
  ``1/2 BPS geometries of M2 giant gravitons,''
  arXiv:hep-th/0504098.
  %%CITATION = HEP-TH 0504098;%%

%\cite{deMelloKoch:2005rq}
\bibitem{deMelloKoch:2005rq}
  R.~de Mello Koch, A.~Jevicki and S.~Ramgoolam,
  ``On exponential corrections to the 1/N expansion in two-dimensional Yang
  Mills theory,''
  arXiv:hep-th/0504115.
  %%CITATION = HEP-TH 0504115;%%

%\cite{Grant:2005qc}
\bibitem{Grant:2005qc}
  L.~Grant, L.~Maoz, J.~Marsano, K.~Papadodimas and V.~S.~Rychkov,
  ``Minisuperspace quantization of 'bubbling AdS' and free fermion droplets,''
  arXiv:hep-th/0505079.
  %%CITATION = HEP-TH 0505079;%%

%\cite{Balasubramanian:2005kk}
\bibitem{Balasubramanian:2005kk}
  V.~Balasubramanian, V.~Jejjala and J.~Simon,
  ``The library of Babel,''
  arXiv:hep-th/0505123.
  %%CITATION = HEP-TH 0505123;%%

%\cite{Ghodsi:2005ks}
\bibitem{Ghodsi:2005ks}
  A.~Ghodsi, A.~E.~Mosaffa, O.~Saremi and M.~M.~Sheikh-Jabbari,
  ``LLL vs. LLM: Half BPS sector of N = 4 SYM equals to quantum Hall system,''
  arXiv:hep-th/0505129.
  %%CITATION = HEP-TH 0505129;%%

%\cite{Mukhi:2005cv}
\bibitem{Mukhi:2005cv}
  S.~Mukhi and M.~Smedback,
  ``Bubbling orientifolds,''
  arXiv:hep-th/0506059.
  %%CITATION = HEP-TH 0506059;%%

%\cite{Boni:2005sf}
\bibitem{Boni:2005sf}
  M.~Boni and P.~J.~Silva,
  ``Revisiting the D1/D5 system or bubbling in AdS(3),''
  arXiv:hep-th/0506085. See also: 
  %%CITATION = HEP-TH 0506085;%%
%\bibitem{Caldarelli:2004ig}
  M.~M.~Caldarelli and P.~J.~Silva,
 ``Giant gravitons in AdS/CFT. I: Matrix model and back reaction,''
   JHEP {\bf 0408}, 029 (2004)
   [arXiv:hep-th/0406096];
   %%CITATION = HEP-TH 0406096;%%
%\bibitem{Caldarelli:2004mz}
  M.~M.~Caldarelli, D.~Klemm and P.~J.~Silva,
  %``Chronology protection in anti-de Sitter,''
  Class.\ Quant.\ Grav.\  {\bf 22}, 3461 (2005)
  [arXiv:hep-th/0411203].
  %%CITATION = HEP-TH 0411203;%%

%\cite{Milanesi:2005tp}
\bibitem{Milanesi:2005tp}
  G.~Milanesi and M.~O'Loughlin,
  ``Singularities and closed time-like curves in type IIB 1/2 BPS geometries,''
  arXiv:hep-th/0507056.
  %%CITATION = HEP-TH 0507056;%%

%\cite{Takayama:2005yq}
\bibitem{Takayama:2005yq}
  Y.~Takayama and A.~Tsuchiya,
  ``Complex Matrix Model and Fermion Phase Space for Bubbling AdS Geometries,''
  arXiv:hep-th/0507070.
  %%CITATION = HEP-TH 0507070;%%

\bibitem{Rastelli}
L. Rastelli, M. Wijnholt. PUPT-2169, Jul 2005. 
e-Print Archive: hep-th/0507037

\bibitem{MarchesiniYQ}
  G.~Marchesini and E.~Onofri,
  ``Planar Limit For SU(N) Symmetric Quantum Dynamical Systems,''
  J.\ Math.\ Phys.\  {\bf 21}, 1103 (1980).
  %%CITATION = JMAPA,21,1103;%%

\bibitem{Halpern:1981fj}
  M.~B.~Halpern and C.~Schwartz,
  ``Large N Classical Solution For The One Matrix Model,''
  Phys.\ Rev.\ D {\bf 24}, 2146 (1981).
  %%CITATION = PHRVA,D24,2146;%%


\bibitem{JY}
A. Jevicki and T. Yoneya, NSF-ITP-93-67, BROWN-HEP-904, UT-KOMABA-93-10, May 1993.
Published in Nucl.Phys.B411:64-96,1994
e-Print Archive: hep-th/9305109

\bibitem{AJnp}
 A.Jevicki
''Nonperturbative Collective Field Theory''
 Nucl.Phys.B388:700-714,1992

\bibitem{IKS}
Satoshi Iso , Dimitria Karabali, B. Sakita ,. CCNY-HEP-92-1, Jan 1992. 18pp.
 Nucl.Phys.B388:700-714,1992
e-Print Archive: hep-th/9202012

\bibitem{JevickiMB}
  A.~Jevicki and B.~Sakita,
  ``The Quantum Collective Field Method And Its Application To The Planar
  Limit,''
  Nucl.\ Phys.\ B {\bf 165}, 511 (1980).
  %%CITATION = NUPHA,B165,511;%%

\bibitem{DasKA}
  S.~R.~Das and A.~Jevicki,
  ``String Field Theory And Physical Interpretation Of D = 1 Strings,''
  Mod.\ Phys.\ Lett.\ A {\bf 5}, 1639 (1990).
  %%CITATION = MPLAE,A5,1639;%%

\bibitem{DemeterfiCW}
  K.~Demeterfi, A.~Jevicki and J.~P.~Rodrigues,
  ``Perturbative results of collective string field theory,''
  Mod.\ Phys.\ Lett.\ A {\bf 6}, 3199 (1991).
  %%CITATION = MPLAE,A6,3199;%%

\bibitem{JevickiYK}
  A.~Jevicki and J.~P.~Rodrigues,
  ``Supersymmetric collective field theory,''
  Phys.\ Lett.\ B {\bf 268}, 53 (1991).
  %%CITATION = PHLTA,B268,53;

\bibitem{RodriguesBY}
  J.~P.~Rodrigues and A.~J.~van Tonder,
  ``Marinari-Parisi and supersymmetric collective field theory,''
  Int.\ J.\ Mod.\ Phys.\ A {\bf 8}, 2517 (1993)
  [arXiv:hep-th/9204061].
  %%CITATION = HEP-TH 9204061;%%

\bibitem{vanTonderVC}
  A.~J.~van Tonder,
  ``A Continuum description of superCalogero models,''
  arXiv:hep-th/9204034.
  %%CITATION = HEP-TH 9204034;%%

\bibitem{Alastair}
Alastair Paulin-Campbell,
``{\it Stabilisation and Lower Dimensional Matrix String Models,}"
(Witwatersrand U.) MSc Dissertation, Nov 1997.

\bibitem{BrezinSV}
  E.~Brezin, C.~Itzykson, G.~Parisi and J.~B.~Zuber,
  ``Planar Diagrams,''
  Commun.\ Math.\ Phys.\  {\bf 59}, 35 (1978).
  %%CITATION = CMPHA,59,35;%%


\bibitem{AvJe}
By Jean Avan, Antal Jevicki (Brown U.),. BROWN-HET-1019, PAR-LPTHE-95-48, Dec 1995. 18pp.
Published in Nucl.Phys.B469:287-301,1996
e-Print Archive: hep-th/9512147

\bibitem{deMelloKochNQ}
  R.~de Mello Koch, A.~Jevicki and J.~P.~Rodrigues,
  ``Collective string field theory of matrix models in the BMN limit,''
  Int.\ J.\ Mod.\ Phys.\ A {\bf 19}, 1747 (2004)
  [arXiv:hep-th/0209155].
  %%CITATION = HEP-TH 0209155;%%

\bibitem{GK}
D.Gross and I.Klebanov,Nucl.Phys.B354:459-474,1991

\bibitem{Mls}
J.Maldacena,Mar 2005. 40pp.
e-Print Archive: hep-th/0503112


\bibitem{AffleckRG}
  I.~Affleck,
  ``Glueballs, Mesons, Surface Roughening And The Collective Field Method.
  (Brown Workshop, Providence 1980, Eds. A. Jevicki and C-I Tan),''
%\href{http://www.slac.stanford.edu/spires/find/hep/www?irn=931039}{SPIRES entry}

\bibitem{RodriguesFS}
  J.~A.~P.~Rodrigues,
  ``Loop, Space, Master Variables And The Spectrum In The Large N Limit,''
   PhD Thesis, Brown Unviversity, 1983, UMI 83-26029
%\href{http://www.slac.stanford.edu/spires/find/hep/www?r=umi\%2F83-26029}{SPIRES entry}


\bibitem{JevickiHB}
  A.~Jevicki and J.~P.~Rodrigues,
  ``Master Variables And Spectrum Equations In Large N Theories,''
  Nucl.\ Phys.\ B {\bf 230}, 317 (1984).
  %%CITATION = NUPHA,B230,317;%%

%\cite{Dobashi:2004nm}
\bibitem{Dobashi:2004nm}
  S.~Dobashi and T.~Yoneya,
  %``Resolving the holography in the plane-wave limit of AdS/CFT
  %correspondence,''
  Nucl.\ Phys.\ B {\bf 711}, 3 (2005)
  [arXiv:hep-th/0406225];
  %%CITATION = HEP-TH 0406225;%% 
%\cite{Dobashi:2004ka}
%\bibitem{Dobashi:2004ka}
  S.~Dobashi and T.~Yoneya,
  ``Impurity non-preserving 3-point correlators of BMN operators from pp-wave
  holography. I: Bosonic excitations,''
  Nucl.\ Phys.\ B {\bf 711}, 54 (2005)
  [arXiv:hep-th/0409058].
  %%CITATION = HEP-TH 0409058;%%

\bibitem{deMelloKochPV}
  R.~de Mello Koch, A.~Donos, A.~Jevicki and J.~P.~Rodrigues,
  ``Derivation of string field theory from the large N BMN limit,''
  Phys.\ Rev.\ D {\bf 68}, 065012 (2003)
  [arXiv:hep-th/0305042].
  %%CITATION = HEP-TH 0305042;%%


\end{thebibliography}
\end{document}